\documentclass[11pt]{article}
\pdfoutput=1

\usepackage{fullpage}
\usepackage{amsmath}
\usepackage{amssymb}
\usepackage{graphicx} 

\usepackage{fancybox,graphicx}
\usepackage[T1]{fontenc}
\usepackage[latin1]{inputenc}
\usepackage{float}
\usepackage{subfigure}
\usepackage{epsfig}

\floatstyle{ruled}
\newfloat{alg}{tbp}{loa}
\floatname{alg}{Algorithm}

\newenvironment{algorithm}[1][\  ] %
{
\rm
\begin{tabbing}
....\=...\=...\=...\=...\=  \+ \kill
} %
{\end{tabbing}
}

\newtheorem{definition}{Definition}[section]
\newtheorem{lemma}{Lemma}[section]
\newtheorem{corollary}{Corollary}[section]
\newtheorem{theorem}{Theorem}[section]
\newtheorem{proposition}{Proposition}[section]
\newtheorem{assumption}{Assumption}[section]

\newcommand{\bx}{{\mathbf x}}

\newcommand{\by}{{\mathbf y}}

\newcommand{\be}{{\mathbf e}}

\newcommand{\rE}{{\mathbb E}}

\newcommand{\cI}{{\cal I}}

\newcommand{\cB}{{\cal B}}
\newcommand{\br}{{\mathbf r}}
\newcommand{\pr}{\mathrm{Pr}}

\newcommand{\Fr}{{\mathrm{supp}}}

\newcommand{\hb}{{\hat{\beta}}}
\newcommand{\bb}{{\bar{\beta}}}
\newcommand{\hQ}{{\hat{Q}}}

\newcommand{\cl}{{\mathrm {cl}}}
\newcommand{\R}{{\mathbb{R}}}
\newcommand{\BlackBox}{\rule{1.5ex}{1.5ex}}
\newenvironment{proof}{\par\noindent{\bf Proof\ }}{\hfill\BlackBox\\[2mm]}

\begin{document}

\title{Learning with Structured Sparsity}
\author{Junzhou Huang \\ Department of Computer Science, Rutgers University
\and Tong Zhang \\ Department of Statistics, Rutgers University
\and Dimitris Metaxas \\ Department of Computer Science, Rutgers
University}
\date{}
\maketitle

\begin{abstract}
This paper investigates a new learning formulation
called  {\em structured sparsity}, which is
a natural extension of the standard sparsity concept in statistical
learning and compressive sensing.
By allowing arbitrary structures on the feature set, this concept
generalizes the group sparsity idea that has
become popular in recent years.
A general theory is developed for learning with
structured sparsity, based on
the notion of coding complexity associated with the structure.
It is shown that if the coding complexity of the target signal
is small, then one can achieve improved performance
by using coding complexity regularization methods,
which generalize the standard sparse regularization.
Moreover, a structured greedy algorithm is proposed to efficiently solve
the structured sparsity problem. It is shown that
the greedy algorithm approximately solves the coding complexity optimization
problem under appropriate conditions.
Experiments are included to demonstrate the advantage of
structured sparsity over standard sparsity on some real applications.
\end{abstract}

\section{Introduction}

We are interested in the sparse learning problem under the fixed design
condition.
Consider a fixed set of $p$ basis vectors $\{\bx_1,\ldots,\bx_p\}$
where $\bx_j \in \R^n$ for each $j$. Here, $n$ is the sample size.
Denote by $X$ the $n \times p$ data matrix, with column $j$
of $X$ being $\bx_j$.
Given a random observation $\by=[\by_1,\ldots,\by_n] \in \R^n$ that depends on
an underlying coefficient vector $\bb \in \R^p$, we are interested
in the problem of estimating $\bb$
under the assumption that the target coefficient $\bb$ is sparse.
Throughout the paper, we consider fixed design only. That is,
we assume $X$ is fixed, and randomization is with respect
to the noise in the observation $\by$.

We consider the situation that $\rE \by$ can be approximated by a
sparse linear combination of the basis vectors:
\[
\rE \by \approx X \bb ,
\]
where we assume that $\bb$ is sparse.
Define the support of a vector $\beta \in \R^p$ as
\[
\Fr(\beta)=\{j: \beta_j \neq 0\} ,
\]
and $\|\beta\|_0 = |\Fr(\beta)|$. A natural method for sparse learning
is $L_0$ regularization:
\[
\hb_{L0}= \arg\min_{\beta \in \R^p}
\hQ(\beta) \quad \text{ subject to } \|\beta\|_0 \leq s ,
\]
where $s$ is the desired sparsity. For simplicity, unless
otherwise stated,
the objective function considered throughout this paper is the
least squares loss
\[
\hQ(\beta)= \|X \beta - \by\|_2^2 ,
\]
although other objective functions for generalized linear models
(such as logistic regression) can be similarly analyzed.

%

Since this
optimization problem is generally NP-hard, in practice, one often
considers approximate solutions.
A standard approach is convex
relaxation of $L_0$ regularization to $L_1$ regularization, often
referred to as Lasso \cite{Tibshirani96}. Another commonly used
approach is greedy algorithms, such as the orthogonal matching
pursuit (OMP) \cite{Tropp07tit}.

In practical applications, one often knows a structure on the
coefficient vector $\bar{\beta}$ in addition to sparsity. For
example, in group sparsity, one assumes that variables in the same
group tend to be zero or nonzero simultaneously. The purpose of
this paper is to study the more general estimation problem under
structured sparsity. If meaningful structures exist, we show that
one can take advantage of such structures to improve the standard
sparse learning.

\section{Related Work}

The idea of using structure in addition to sparsity has been
explored before.
An example is group structure, which has received much attention
recently. For example, group sparsity has been
considered for simultaneous sparse approximation \cite{Wipf07TSP}
and multi-task compressive sensing \cite{Ji08TSP} from the
Bayesian hierarchical modeling point of view. Under the Bayesian
hierarchical model framework, data from all sources contribute to
the estimation of hyper-parameters in the sparse prior model. The
shared prior can then be inferred from multiple sources. He et al.
recently extend the idea to the tree sparsity in the Bayesian
framework \cite{He08, He09}. Although the idea can be justified
using standard Bayesian intuition, there are no theoretical
results showing how much better (and under what kind of
conditions) the resulting algorithms perform.
In the statistical literature, Lasso has been extended to the
group Lasso when there exist group/block structured dependences
among the sparse coefficients \cite{Yuan06JRSS}.

However, none of the above mentioned work was able to show advantage
of using group structure. Although some
theoretical results were developed in
\cite{Bach08-groupLasso,NarRin08}, neither showed that group Lasso
is superior to the standard Lasso. The authors of \cite{KolYua08}
showed that group Lasso can be superior to standard Lasso when each
group is an infinite dimensional kernel, by relying on the fact
that meaningful analysis can be obtained for kernel methods in
infinite dimension. In \cite{ObWaJo08}, the authors consider a
special case of group Lasso in the multi-task learning scenario,
and show that the number of samples required for recovering the
exact support set is smaller for group Lasso under appropriate
conditions.
In \cite{HuangZhang09:group}, a theory for
group Lasso was developed using a concept called strong group sparsity,
which is a special case of the general structured sparsity idea
considered here. It was shown in \cite{HuangZhang09:group}
that group Lasso is superior to standard Lasso for strongly
group-sparse signals, which provides a convincing theoretical
justification for using group structured sparsity.

While group Lasso works under the strong group sparsity assumption,
it doesn't handle the more general structures considered in this paper.
Several limitations of group Lasso were mentioned in
\cite{HuangZhang09:group}.
For example, group Lasso does not correctly handle overlapping groups
(in that overlapping components are over-counted);
that is, a given coefficient should not belong to different groups.
This requirement is too rigid for many practical applications.
To address this issue, a method called
composite absolute penalty (CAP) is proposed
in \cite{Zhao09AS} which can handle
overlapping groups. Unfortunately, no theory is established
to demonstrate the effectiveness of the approach.
In a related development \cite{Kowalski09spasr},
Kowalski et al. generalized the mixed norm penalty to structured
shrinkage, which can identify the structured significance maps and
thus can handle the case of the overlapping groups, However, the
structured shrinkage operations do not necessarily
convergence to a fixed point. There were no additional theory to
justify their methods.



Other structures have also been explored in the literature.
For example, so-called tonal and transient structures were considered for
sparse decomposition of audio signals in \cite{Daudet04}, but again without
any theory. Grimm et al. \cite{Grimm07Arch} investigated
positive polynomials with structured sparsity from an optimization
perspective. The theoretical result there did not address
the effectiveness of such methods in comparison to standard sparsity.
The closest work to ours is a recent paper \cite{Baraniuk08model},
which we learned after finishing this paper.
In that paper, a specific case of structured sparsity,
referred to as model based sparsity, was considered.
It is important to note that some theoretical results were obtained there
to show the effectiveness of their method in compressive sensing, although
in a more limited scope than results presented here.
Moreover, they do not provide a generic framework for structured sparsity.
In their algorithm, different schemes have to be specifically designed for
different data models, and under specialized assumptions.
It remains as an open issue how to develop a
general theory for structured sparsity, together with a general algorithm
that can be applied to a wide class of such problems.


We see from the above discussion that there exists extensive literature
on structured sparsity, with empirical evidence showing that
one can achieve better performance by imposing additional structures.
However, none of the previous work was able to establish a general theoretical
framework for structured sparsity that can quantify
its effectiveness.
The goal of this paper is to develop such a general theory
that addresses the following issues, where we pay special attention
to the benefit of  structured sparsity over the standard
non-structured sparsity:
\begin{itemize}
\item quantifying structured sparsity;
\item the minimal number of measurements required in
  compressive sensing;
\item estimation accuracy under stochastic noise;
\item an efficient algorithm that can solve a wide class of
  structured sparsity problems.
\end{itemize}

\section{Structured Sparsity}

In structured sparsity, not all sparse patterns are equally likely. For example, in group sparsity, coefficients within the same group are more likely to be zeros or nonzeros simultaneously. This means that if a sparse coefficient vector's support set is consistent with the underlying group structure, then it is more likely to occur, and hence incurs a smaller penalty in learning. One contribution of this work is to formulate how to define structure on top of sparsity, and how to penalize each sparsity pattern.

In order to formalize the idea, we denote by $\cI=\{1,\ldots,p\}$ the index set of the coefficients.
Consider any sparse subset $F \subset \{1,\ldots,p\}$, we assign a
cost $\cl(F)$.
In structured sparsity, the cost of $F$ is an upper bound of
the coding length of $F$ (number of bits needed to represent $F$
by a computer program) in a pre-chosen prefix coding scheme. It is a well-known
fact in information theory (e.g. \cite{CoverThomas91})
that mathematically, the existence of such a
coding scheme is equivalent to
\[
\sum_{F \subset \cI} 2^{-\cl(F)} \leq 1.
\]
From the Bayesian statistics point of view,
$2^{-\cl(F)}$ can be regarded as a lower bound of the probability of $F$.
The probability model of structured sparse learning is thus:
first generate the sparsity pattern $F$ according to probability $2^{-\cl(F)}$;
then generate the coefficients in $F$.

\begin{definition} \label{def:sparsity}
  A cost function $\cl(F)$ defined on subsets of $\cI$ is called a
  coding length (in base-2) if
  \[
  \sum_{F \subset \cI, F \neq \emptyset} 2^{-\cl(F)} \leq 1 .
  \]
  We give $\emptyset$ a coding length 0.
  The corresponding structured sparse
  coding complexity of $F$ is defined as
  \[
  c(F) = |F| + \cl(F) .
  \]
  A coding length $\cl(F)$ is sub-additive if
  \[
  \cl(F \cup F') \leq \cl(F) + \cl(F') ,
  \]
  and a coding complexity $c(F)$ is sub-additive if
  \[
  c(F \cup F') \leq c(F) + c(F') .
  \]
\end{definition}

Clearly if $\cl(F)$ is sub-additive, then the corresponding coding
complexity $c(F)$ is also sub-additive.
Based on the structured coding complexity of subsets of $\cI$,
we can now define the structured coding complexity of a sparse
coefficient vector $\bb \in \R^p$.
\begin{definition}
  Giving a coding complexity $c(F)$,
  the structured sparse coding complexity of
  a coefficient vector $\bb \in \R^p$ is
  \[
  c(\bb)=\min \{ c(F) : \Fr(\bb) \subset F \} .
  \]
\end{definition}

Later in the paper, we will show that if a coefficient vector $\bb$ has a small
coding complexity $c(\bb)$, then
$\bb$ can be effectively learned, with good in-sample prediction performance
(in statistical learning) and reconstruction performance (in compressive sensing).
In order to see why the definition requires adding $|F|$ to $\cl(F)$,
we consider the generative model for structured sparsity mentioned earlier.
In this model, the number of bits
to encode a sparse coefficient vector is the sum of the number of bits to
encode $F$ (which is $\cl(F)$) and the number of
bits to encode nonzero coefficients
in $F$ (this requires $O(|F|)$ bits up to a fixed precision).
Therefore the total number of bits
required is $\cl(F) + O(|F|)$.
This information theoretical result translates into a statistical estimation
result: without additional
regularization, the learning complexity for least squares regression
within any fixed support set $F$ is $O(|F|)$. By adding the model selection
complexity $\cl(F)$ for each support set $F$, we obtain an overall
statistical estimation complexity of $O(\cl(F)+|F|)$.

While the idea of using coding based penalization
is clearly motivated by the minimum description length (MDL) principle,
the actual penalty we obtain for structured sparsity problems is different from
the standard MDL penalty for model selection. This difference is important
in sparse learning.
Therefore in order to prevent confusion, we avoid using MDL in our
terminology.
Nevertheless, one may consider our framework as a natural combination of the
MDL idea and the modern sparsity analysis.

\section{Structured Sparsity Examples}
\label{sec:examples}

Before giving detailed examples, we introduce a general coding scheme
called {\em block coding}.
The basic idea of block coding
is to define a coding scheme on a small number of base blocks
(a block is a subset of $\cI$),
and then define a coding scheme on all subsets of $\cI$ using these
base blocks.

Consider a subset $\cB \subset 2^\cI$. That is, each element (a block)
of $\cB$ is a subset of $\cI$.
We call $\cB$ a block set if
$\cI = \cup_{B \in \cB} B$ and all single element
sets $\{j\}$ belong to $\cB$ ($j \in \cI$). Note that $\cB$ may contain
additional non single-element blocks.
The requirement of $\cB$ containing all single element sets is for convenience,
as it implies that every subset $F \subset \cI$ can be expressed as the
union of blocks in $\cB$.

Let $\cl_0$ be a code length on $\cB$:
\[
\sum_{B \in \cB} 2^{-\cl_0(B)} \leq 1 ,
\]
we define $\cl(B)=\cl_0(B)+1$ for $B \in \cB$.
It not difficult to show that the following cost function on
$F \subset \cI$ is a coding length
\[
\cl(F)= \min \left\{ \sum_{j=1}^b \cl(B_j)
  : F = \bigcup_{j=1}^b B_j \quad (B_j \in \cB) \right\} .
\]
This is because
\[
\sum_{F \subset \cI, F \neq \emptyset}  2^{-\cl(F)}
\leq \sum_{b \geq 1} \sum_{\{B_\ell\} \in \cB^b} 2^{-\sum_{\ell=1}^b \cl(B_\ell)}
\leq \sum_{b \geq 1} \prod_{\ell=1}^b \sum_{B_\ell \in \cB} 2^{-\cl(B_\ell)}
\leq \sum_{b \geq 1} 2^{-b} = 1 .
\]
It is clear from the definition that block coding is sub-additive.

The main purpose of introducing block coding is to design computationally
efficient algorithms based on the block structure.
In particular, this paper considers a
structured greedy algorithm that can take advantage of block structures.
In the structured greedy algorithm, instead of searching over all subsets of
$\cI$ up to a fixed coding complexity $s$ (the number of such subsets can be exponential in $s$), we greedily add blocks from $\cB$ one at a time.
Each search problem over $\cB$ can be efficiently performed because
we require that $\cB$ contains only a computationally manageable number
of base blocks. Therefore the algorithm is computationally efficient.

We will show that under appropriate conditions,
a target coefficient vector with a small block coding complexity
can be approximately learned using the structured greedy algorithm.
This means that the block coding scheme has important algorithmic implications.
That is, if a coding scheme can be approximated by block coding with
a small number of base blocks,
then the corresponding estimation problem can be approximately
solved using the structured greedy algorithm.
For this reason, we shall pay special attention to block coding approximation
schemes for examples discussed below.

\subsection*{Standard sparsity}

A simple coding scheme is to code each subset $F \subset \cI$ of cardinality
$k$ using $k\log_2 (2p)$ bits, which corresponds to block coding
with $\cB$ consisted only of single element sets, and each base block
has a coding length $\cl_0=\log_2 p$.
This corresponds to the complexity for the standard sparse learning.

A more general version is to consider single element blocks $\cB=\{ \{j\}: j \in \cI \}$, with a non-uniform coding scheme $\cl_0(\{j\})= c_j$, such that
$\sum_j 2^{-c_j} \leq 1$. It leads to a non-uniform coding length on $\cI$ as
\[
\cl(B)= |B| + \sum_{j \in B} c_j .
\]
In particular, if a feature $j$ is likely to be nonzero, we should give it
a smaller coding length $c_j$, and if a feature $j$ is likely to be zero,
we should give it a larger coding length.

\subsection*{Group sparsity}

The concept of group sparsity has appeared in various recent work,
such as the group Lasso in \cite{Yuan06JRSS}.
Consider a partition of $\cI=\cup_{j=1}^m G_j$ into $m$ disjoint groups.
Let $\cB_G$ contain the $m$ groups $\{G_j\}$, and $\cB_1$ contain
$p$ single element blocks.
The strong group sparsity coding scheme
is to give each element in $\cB_1$ a code-length $\cl_0$
of $\infty$, and each element in $\cB_G$ a code-length $\cl_0$ of $\log_2 m$.
Then the block coding scheme with blocks $\cB=\cB_G \cup \cB_1$ leads to
group sparsity, which only looks for signals consisted of the groups.
The resulting coding length is:
$\cl(B) = g \log_2 (2m)$ if $B$ can be represented as the union of $g$ disjoint
groups $G_j$; and $\cl(B)=\infty$ otherwise.

Note that if the signal can be expressed as
the union of $g$ groups, and each group size is
$k_0$, then the group coding length $g \log_2(2m)$ can be significantly
smaller than the standard sparsity coding length of $g k_0 \log_2(p)$.
As we shall see later, the smaller coding complexity implies better
learning behavior, which
is essentially the advantage of using group sparse structure.
It was shown in \cite{HuangZhang09:group} that strong
group sparsity defined above also characterizes
the performance group Lasso. Therefore if a signal has a pre-determined
group structure, then group Lasso is superior to standard Lasso.

An extension of this idea is to allow more general block coding length
for $\cl_0(G_j)$ and $\cl_0(\{j\})$ so that
\[
\sum_{j=1}^m 2^{-\cl_0(G_j)} + \sum_{j=1}^p 2^{-\cl_0(\{j\})} \leq 1 .
\]
This leads to non-uniform coding of the groups, so that a group that is
more likely to be nonzero is given a smaller coding length.

\subsection*{Hierarchical sparsity}

One may also create a hierarchical group structure. A simple
example is wavelet coefficients of a signal \cite{Mallat99}.
Another simple example is a binary tree with the variables as
leaves, which we describe below. Each internal node in the tree is
associated with three options: left child only, right child only,
and both children; each option can be encoded in $\log_2 3$ bits.

Given a subset $F \subset \cI$,
we can go down from the root of the tree, and at each node, decide whether
only left child contains elements of $F$, or only right child contains
elements of $F$, or both children contain elements of $F$.
Therefore the coding length of $F$ is
$\log_2 3$ times the total number of internal nodes
leading to elements of $F$. Since each leaf corresponds to no more than
$\log_2 p$ internal nodes, the total coding length is no worse than
$\log_2 3 \log_2 p |F|$. However, the coding length can be significantly smaller if nodes are close to each other or are clustered. In the extreme case, when the nodes are consecutive, we have $O(|F|+\log_2 p)$ coding length.
More generally, if we can order elements in $F$ as $F=\{j_1,\ldots,j_q\}$, then the coding length can be bounded as $\cl(F)=O(|F|+\log_2 p + \sum_{s=2}^q \log_2 \min_{\ell<s} |j_s-j_\ell|)$.

If all internal nodes of the tree are also variables in $\cI$
(for example, in the case of wavelet decomposition), then one may consider feature set $F$ with the
following property: if a node is selected, then its parent is also selected.
This requirement is very effective in wavelet compression, and
often referred to as the zero-tree structure
\cite{Shapiro93}. Similar requirements have also been
applied in statistics \cite{Zhao09AS} for variable selection.
The argument presented in this section shows that if we require $F$
to satisfy the zero-tree structure, then its coding length is at most
$O(|F|)$, without any explicit dependency on the dimensionality $p$.
This is because one does not have to reach a leave node.

The tree-based coding scheme discussed in this section
can be approximated by block coding using no more than
$p^{1+\delta}$ base blocks ($\delta>0$).
The idea is similar to that of the
image coding example in the more general graph
sparsity scheme which we discuss next.

\subsection*{Graph sparsity}

We consider a generalization of the hierarchical and group
sparsity idea that employs a (directed or undirected) graph structure
$G$ on $\cI$. To the best of our knowledge, this general structure
has not been considered in any previous work.

In graph sparsity, each variable (an element of $\cI$) is a node of $G$
but $G$ may also contain additional nodes that are not variables.
For simplicity, we assume $G$ contains a starting node
(this requirement is not critical).

At each node $v \in G$, we define coding length $\cl_v(S)$ on all subsets
$S$ of the neighborhood $N_v$ of $v$ including the empty set,
as well as any other single node $u \in G$ with
$\cl_v(u)$, such that
\[
\sum_{S \subset N_v} 2^{-\cl_v(S)} + \sum_{u \in G} 2^{-\cl_v(u)} \leq 1 .
\]

To encode $F \subset G$, we start with the active set containing only the
starting node, and finish when the set becomes empty.
At each node $v$ before termination, we may either pick a subset
$S \subset N_v$, with coding length $\cl_v(S)$, or a node in $u \in G$,
with coding length $\cl_v(u)$, and then put the selection into the active set.
We then remove $v$ from the active set
(once a node $v$ is removed, it does not return to the active
set anymore). This process is continued until the active set becomes empty.

As a concrete example, we consider image processing, where each
image is a rectangle of pixels (nodes); each pixel
is connected to four adjacent pixels, which forms the underlying graph structure.
At each pixel, the number of subsets in its neighborhood is $2^{4}=16$
(including the empty set), and each subset is given
a coding length $\cl_v(S)=5$;
we also encode all other pixels in the image
with random jumping, each with a coding length $1+ \log_2 p$.
Using this scheme, we can encode each connected region $F$
by no more than $\log_2 p + 5 |F|$ bits by growing the region from a
single point in the region.
Therefore if $F$ is composed
of $g$ connected regions, then the coding length is
$g\log_2 p + 5 |F|$, which can be significantly better than
standard sparse coding length of $|F| \log_2 p$.

This example shows that the general graph coding scheme presented here favors connected regions (that is, nodes that are grouped together with respect to the graph structure). In particular, it proves the following more general
result.
\begin{proposition}
  Given a graph $G$,
  there exists a constant $C_G$ such that
  for any probability distribution $q$ on $G$ ($\sum_{v \in G} q(v)=1$ and
  $q(v) \geq 0$ for $v \in G$), the following quantity
  is a coding length on $2^G$:
  \[
  \cl(F) = C_G |F|  - \sum_{j=1}^g \max_{v \in F_j} \log_2 q(v) ,
  \]
  where $F \subset 2^G$ can be decomposed into
  the union of $g$ connected components $F=\cup_{j=1}^g F_j$.
\end{proposition}
Our simple and suboptimal coding scheme described for the
image processing example gives an upper bound
$C_G \leq 1+d_G$, where $d_G$ is the
maximum degree of $G$. However, for many graphs, one can improve this
constant to $O(\log d_G)$ with a slightly more complicated argument.

As a simple application of graph coding, we consider the special case where
we have only one connected component that contains the starting node $v_0$.
We can simply let $q(v_0)=1$, and the coding length is
$O(|F|)$, which is independent of the dimensionality $p$.
This generalizes the similar claim for
the zero-tree structure described earlier.

The graph coding scheme can be approximated with block coding.
The idea is to consider relatively small sized base blocks consisted of nodes
that are close together with respect to the graph structure, and then
use the induced block coding scheme to approximate the graph coding.

For example, for the previously discussed image coding example, we can
use connected blocks of size upto $\delta \log_2 p/5$ as base blocks
of $\cB$ ($\delta>0$).
Since each base block can be encoded with $(1+\delta) \log_2 p$ bits
by earlier discussion,
we know that the total number of base blocks can be no more than $p^{1+\delta}$.
We can give each of such blocks a coding length $(1+\delta)\log_2 p$.
For a connected region $F$ that can be covered by $O(1+|F|/\log_2 p)$
of such blocks, the corresponding block coding length for a subset $F$ is
$\cl(F)=O(|F| + \log_2 p)$, which is the same as the complexity of
the original graph coding length (up to a constant).
This means that graph coding length can be approximated with block coding
scheme. As we have pointed out, such an approximation is useful because
the latter is required in the structured greedy algorithm which we propose
in this paper.

\subsection*{Random field sparsity}

Let $z_j \in \{0,1\}$ be a random variable for $j \in \cI$ that indicates
whether $j$ is selected or not.
The most general coding scheme is to consider a joint probability distribution
of $z=[z_1,\ldots,z_p]$.
The coding length for $F$ can be defined as $-\log_2 p(z_1,\ldots,z_p)$
with $z_j = I(j \in F)$ indicating whether $j \in F$ or not.

Such a probability distribution can often be conveniently represented as
a binary random field on an underlying graph.
In order to encourage sparsity, on average,
the marginal probability $p(z_j)$ should take 1 with probability close to $O(1/p)$, so that the expected number of $j$'s with $z_j=1$ is $O(1)$.
For disconnected graphs ($z_j$ are independent), the variables $z_j$ are iid Bernoulli random variables with probability $1/p$ being one.
In this case, the coding length
of a set $F$ is $|F|\log_2 (p) - (p-|F|) \log_2(1-1/p) \approx |F| \log_2(p) +1$. This is essentially the probability model for the standard sparsity scheme.

In many cases, it is possible to approximate a general random field
coding scheme with block coding
by using approximation methods in the graphical model literature.
However, the details of such approximations
are beyond the scope of this paper.

\section{Algorithms for Structured Sparsity}
\label{sec:algorithm}
The following algorithm is a natural extension of $L_0$ regularization
to structured sparsity problems. It penalizes the coding complexity
instead of the cardinality  (sparsity) of the feature set.
\begin{equation}
\hb_{constr} = \arg\min_{\beta \in \R^p}
\hQ(\beta)
\quad \text{subject to }  c(\beta) \leq s .
\label{eq:struct-constr}
\end{equation}
Alternatively, we may consider the formulation
\begin{equation}
\hb_{pen} = \arg\min_{\beta \in \R^p} \left[
\hQ(\beta) + \lambda c(\beta)
\right] .
\label{eq:struct-pen}
\end{equation}

The optimization of either (\ref{eq:struct-constr}) or (\ref{eq:struct-pen})
is generally hard. For related problems,
there are two common approaches to alleviate this difficulty.
One is convex relaxation ($L_1$ regularization to replace $L_0$ regularization
for standard sparsity); the other is forward greedy selection
(also called orthogonal matching pursuit or OMP).
We do not know any extensions of $L_1$ regularization like
convex relaxation that can handle general structured sparsity formulations.
However, one can extend greedy algorithm by using a block structure.
We call the resulting procedure structured greedy algorithm or
StructOMP, which can approximately solve (\ref{eq:struct-constr}).

We have discussed the relationship of this greedy algorithm and block coding
in Section~\ref{sec:examples}.
It is important to understand that the block structure is only used to limit
the search space in the greedy algorithm. The actual coding scheme does not
have to be the corresponding block coding. However, our theoretical analysis
assumes that the underlying coding scheme can be approximated with block
coding using base blocks employed in the greedy algorithm.
Although one does not need to know the specific approximation in order to use the greedy algorithm, knowing its existence (which can be shown for the examples discussed in Section~\ref{sec:examples}) guarantees the effectiveness of the algorithm. It is also useful to understand that our result does not imply that the algorithm won't be effective if the actual coding scheme cannot be approximated by block coding.

\begin{figure}[ht]
  \centering
  \begin{Sbox}
    \begin{minipage}{1.0\linewidth}
      \begin{algorithm}
        Input: $(X,\by)$, $\cB \subset 2^\cI$, $s>0$ \\
        Output: $F^{(k)}$ and $\beta^{(k)}$ \\
        let $F^{(0)}=\emptyset$ and $\beta^{(0)}=0$ \\
        {\bf for} $k=1, 2, \ldots $ \+ \\
        select $B^{(k)} \in \cB \text{ to maximize progress} \qquad (*)$ \\
        let $F^{(k)} = B^{(k)} \cup F^{(k-1)}$ \\
        let $\beta^{(k)}= \arg\min_{\beta \in \R^p} \hQ(\beta) \text{ subject to } \Fr(\beta) \subset F^{(k)}$ \\
        {\bf if} ($c(\beta^{(k)}) >s$) {\bf break} \- \\
        {\bf end}
      \end{algorithm}
    \end{minipage}
  \end{Sbox}\fbox{\TheSbox}
  \caption{Structured Greedy Algorithm}
\label{fig:forward-greedy}
\end{figure}

In Figure~\ref{fig:forward-greedy}, we are given a set of blocks
$\cB$ that contains subsets of $\cI$. Instead of searching all subsets
$F \subset \cI$ up to a certain complexity $|F|+c(F)$, which is
computationally infeasible,
we search only the blocks restricted to $\cB$.
It is assumed that searching over $\cB$ is computationally manageable.

At each step $(*)$, we try to find a block from $\cB$ to maximize
progress. It is thus necessary to define a quantity that
measures progress. Our idea is to approximately maximize the gain ratio:
\[
\lambda^{(k)}=\frac{\hQ(\beta^{(k-1)})-\hQ(\beta^{(k)})}{c(\beta^{(k)})-c(\beta^{k-1})} ,
\]
which measures the reduction of objective function per unit increase of
coding complexity.
This greedy criterion is a natural generalization of the standard greedy
algorithm, and essential in our analysis.
For least squares regression, we
can approximate $\lambda^{(k)}$ using the following definition
\begin{equation}
\phi(B)=\frac{\|P_{B-F^{(k-1)}}(X\beta^{(k-1)}-\by)\|_2^2}
{c(B \cup F^{(k-1)})-c(F^{(k-1)})} , \label{eq:phi}
\end{equation}
where
\[
P_F = X_F (X_F^\top X_F)^{-1} X_F^\top
\]
is the projection matrix to the subspaces generated by columns of $X_F$.
We then select $B^{(k)}$ so that
\[
\phi(B^{(k)}) \geq \gamma \max_{B \in \cB}  \phi(B) ,
\]
where $\gamma \in (0,1]$ is a fixed approximation
ratio that specifies the quality of approximate optimization.
Alternatively, we may use a simpler definition
\[
\tilde{\phi}(B)=\frac{\|X_{B-F^{(k-1)}}^\top (X\beta^{(k-1)}-\by)\|_2^2}{c(B \cup F^{(k-1)})-c(F^{(k-1)})} ,
\]
which is easier to compute, especially when blocks are overlapping.
Since the ratio
\[
\|X_{B-F^{(k-1)}}^\top \br\|_2^2/\|P_{B-F^{(k-1)}}\br\|_2^2
\]
is bounded between $\rho_+(B)$ and $\rho_-(B)$
(these quantities are defined in Definition~\ref{def:rho}),
we know that maximizing
$\tilde{\phi}(B)$ would lead to approximate maximization of
$\phi(B)$ with $\gamma \geq \rho_-(B)/\rho_+(B)$.

Note that we shall ignore $B \in \cB$ such that $B \subset F^{(k-1)}$, and just
let the corresponding gain to be $0$. Moreover, if there exists a base block
$B \not\subset F^{(k-1)}$ but
$c(B \cup F^{(k-1)}) \leq c(F^{(k-1)})$, we can always select $B$
and let $F^{(k)}=B \cup F^{(k-1)}$ (this is because it is always beneficial to add more features into $F^{(k)}$ without additional coding complexity).
We assume this step is always performed
if such a $B \in \cB$ exists. The non-trivial case is
$c(B \cup F^{(k-1)}) > c(F^{(k-1)})$ for all $B \in \cB$; in this case
both $\phi(B)$ and $\tilde{\phi}(B)$ are well defined.

\section{Theory of Structured Sparsity}

\subsection{Assumptions}
We assume sub-Gaussian noise as follows.
\begin{assumption}\label{assump:noise}
Assume that $\{\by_i\}_{i=1,\ldots,n}$ are independent (but not necessarily identically distributed) sub-Gaussians:
there exists a constant $\sigma \geq 0$
such that $\forall i$ and $\forall t \in R$,
\[
\rE_{\by_i} \; e^{t (\by_i-\rE {\by_i})} \leq e^{\sigma^2 t^2/2} .
\]
\end{assumption}

Both Gaussian and bounded random variables are sub-Gaussian
using the above definition.
For example, if a random variable
$\xi \in [a,b]$, then $\rE_\xi e^{t (\xi - \rE \xi)} \leq e^{(b-a)^2 t^2/8}$.
If a random variable is Gaussian: $\xi \sim N(0,\sigma^2)$, then
$\rE_\xi e^{t \xi} \leq e^{\sigma^2 t^2/2}$.

The following property of sub-Gaussian noise is important in our analysis.
Our simple proof yields a sub-optimal choice of the constants.
\begin{proposition} \label{prop:noise}
  Let $P \in \R^{n \times n}$ be a projection matrix of rank $k$,
  and $\by$ satisfies Assumption~\ref{assump:noise}.
  Then
  for all $\eta \in (0,1)$, with probability larger than $1-\eta$:
  \[
  \|P (\by -\rE \by)\|_2^2 \leq  \sigma^2 [ 7.4 k + 2.7 \ln (2/\eta) ] .
  \]
\end{proposition}

We also need to generalize sparse eigenvalue condition, used in the
modern sparsity analysis.
It is related to (and weaker than) the RIP (restricted isometry property)
assumption \cite{CandTao05-rip} in the compressive sensing literature.
This definition takes advantage of coding complexity, and can be
also considered as (a weaker version of) structured RIP.
We introduce a definition.
\begin{definition} \label{def:rho}
For all $F \subset \{1,\ldots,p\}$, define
\begin{align*}
\rho_-(F)=&\inf \left\{\frac{1}{n}\|X \beta\|_2^2/\|\beta\|_2^2: \Fr(\beta) \subset F\right\} , \\
  \rho_+(F)=&\sup \left\{\frac{1}{n}\|X \beta\|_2^2/\|\beta\|_2^2: \Fr(\beta) \subset F\right\} .
\end{align*}
Moreover, for all $s>0$, define
  \begin{align*}
    \rho_-(s) =& \inf \{\rho_-(F): F \subset \cI, c(F) \leq s \} , \\
    \rho_+(s) =& \sup \{\rho_+(F): F \subset \cI, c(F) \leq s \} .
  \end{align*}
\end{definition}

In the theoretical analysis, we need to assume that
$\rho_-(s)$ is not too small for some $s$ that is larger than
the signal complexity.
Since we only consider eigenvalues for submatrices with
small cost $c(\bb)$,
the sparse eigenvalue $\rho_-(s)$ can be significantly larger than
the corresponding ratio for standard sparsity
(which will consider all subsets of $\{1,\ldots,p\}$
up to size $s$).
For example, for
random projections used in compressive sensing applications,
the coding length $c(\Fr(\bb))$ is $O(k \ln p)$ in standard sparsity,
but can be as low as $c(\Fr(\bb))=O(k)$ in structured sparsity (if we can guess
$\Fr(\bb)$ approximately correctly.
Therefore instead of requiring
$n= O(k \ln p)$ samples, we requires only $O(k + \cl(\Fr(\bb)))$.
The difference can be significant when
$p$ is large and the coding length $\cl(\Fr(\bb)) \ll k \ln p$.
An example for this is group sparsity, where we have $p/k_0$ even sized
groups, and variables in each group are simultaneously zero or nonzero.
The coding length of the groups are $(k/k_0) \ln (p/k_0)$, which is significantly smaller than $k \ln p$ when $p$ is large.

More precisely, we have the following random projection sample complexity
bound for the structured sparse eigenvalue condition. The theorem implies
that the structured RIP condition is
satisfied with sample size $n=O((k/k_0) \ln (p/k_0))$ in group sparsity
rather than $n=O(k \ln (p))$ in standard sparsity. Therefore Theorem~\ref{thm:struct-constr} shows that in the compressive sensing applications, it is possible to reconstruct signals with fewer number of random projections by using group sparsity (or more general structured sparsity).

\begin{theorem} [Structured-RIP]\label{thm:rip}
Suppose that elements in $X$ are iid standard Gaussian random variables
$N(0,1)$.
For any $t>0$ and $\delta \in (0,1)$, let
\[
n \geq \frac{8}{\delta^2}
[ \ln 3 + t  + s \ln (1+ 8/\delta) ] .
\]
Then with probability at least $1-e^{-t}$,
the random matrix $X \in \mathbb{R}^{n \times p}$
satisfies the following structured-RIP inequality
for all vector $\bb \in \R^p$ with coding complexity no more than $s$:
\begin{equation}
 (1 -\delta)\| \bb \|_{2} \leq
\frac{1}{\sqrt{n}}
\| X \bb \|_{2} \leq (1+\delta)\| \bb \|_{2}. \label{eq:Group-RIP}
\end{equation}
\end{theorem}

Although in the theorem,
we assume Gaussian random matrix in order to state explicit
constants, it is clear that
similar results hold for other sub-Gaussian random matrices.

\subsection{Coding complexity regularization}
The following result gives a performance bound for
constrained coding complexity regularization.
\begin{theorem} \label{thm:struct-constr}
Suppose that Assumption~\ref{assump:noise} is valid.
Consider any fixed target $\bb \in \R^p$.
Then with probability exceeding $1-\eta$,
for all $\epsilon \geq 0$ and $\hb \in \R^p$ such that:
$\hQ(\hb) \leq \hQ(\bb) + \epsilon$,
we have
\[
\|X \hb- \rE \by \|_2 \leq \|X \bb - \rE \by\|_2
+ \sigma \sqrt{2 \ln(6/\eta)}
+ 2  (7.4 \sigma^2 c(\hb) + 4.7 \sigma^2 \ln (6/\eta) + \epsilon )^{1/2} .
\]
Moreover, if the coding scheme $c(\cdot)$ is sub-additive, then
\[
n \rho_-(c(\hb)+c(\bb)) \|\hb- \bb \|_2^2
\leq 10 \|X \bb- \rE\by\|_2^2 + 37 \sigma^2  c(\hb)
+ 29 \sigma^2 \ln (6/\eta) + 2.5 \epsilon .
\]
\end{theorem}

This theorem immediately implies the following result
for (\ref{eq:struct-constr}):
$\forall \bb$ such that $c(\bb) \leq s$,
\begin{align*}
&\frac{1}{\sqrt{n}}\|X \hb_{constr} - \rE \by \|_2 \leq \frac{1}{\sqrt{n}}
\|X \bb - \rE \by\|_2
+ \frac{\sigma}{\sqrt{n}} \sqrt{2\ln(6/\eta)}
+ \frac{2\sigma}{\sqrt{n}} (7.4 s + 4.7 \ln (6/\eta))^{1/2} , \\
& \|\hb_{constr}- \bb \|_2^2 \leq
\frac{1}{\rho_-(s+c(\bb)) n}
\left[ 10 \|X \bb- \rE\by\|_2^2 + 37 \sigma^2  s + 29 \sigma^2 \ln (6/\eta)
\right] .
\end{align*}
Note that we generally expect $\rho_-(s+c(\bb))=O(1)$.
The result immediately implies that as sample size $n \to \infty$
and $s/n \to 0$,
the root mean squared error prediction performance
$\|X \hb- \rE \by \|_2/\sqrt{n}$ converges to the optimal prediction performance
$\inf_{c(\bb)\leq s} \|X \bb- \rE \by \|_2/\sqrt{n}$.
This result is agnostic in that
even if $ \|X \bb- \rE \by \|_2/\sqrt{n}$ is large, the result is still meaningful
because it says the performance of the estimator $\hb$ is competitive to
the best possible estimator in the class $c(\bb) \leq s$.

In compressive sensing applications, we take $\sigma=0$, and we are interested
in recovering $\bb$ from random projections. For simplicity,
we let $X \bb = \rE \by = \by$, and our result shows that the constrained
coding complexity penalization method achieves exact
reconstruction $\hb_{constr}= \bb$ as long as $\rho_-(2c(\bb)) >0$
(by setting $s=c(\bb)$).
According to Theorem~\ref{thm:rip}, this is possible when
the number of random projections (sample size)
reaches $n=O(c(\bb))$. This is a generalization of corresponding
results in compressive sensing \cite{CandTao05-rip}.
As we have pointed out earlier, this number can be
significantly smaller than the standard sparsity requirement of
$n=O(\|\bb\|_0 \ln p)$, if the structure imposed in the formulation
is meaningful.

Similar to Theorem~\ref{thm:struct-constr},
we can obtain  the following result for (\ref{eq:struct-pen}).
A related result for standard sparsity under Gaussian noise can be
found in \cite{BuTsWe07-aos}.
\begin{theorem} \label{thm:struct-pen}
Suppose that Assumption~\ref{assump:noise} is valid.
Consider any fixed target $\bb \in \R^p$.
Then with probability exceeding $1-\eta$,
for all $\lambda > 7.4 \sigma^2$ and
$a \geq 7.4\sigma^2 /(\lambda- 7.4\sigma^2)$, we have
\[
\|X \hb_{pen} - \rE \by\|_2^2 \leq (1+a)^2 \|X \bb - \rE \by \|_2^2
 + (1+a) \lambda c(\bb)  +
\sigma^2 (10 + 5 a+ 7 a^{-1})\ln (6/\eta) .
\]
\end{theorem}
Unlike the result for  (\ref{eq:struct-constr}), the prediction
performance $\|X \hb_{pen} - \rE \by\|_2$ of the estimator in
(\ref{eq:struct-pen}) is competitive to
$(1+a)\|X \bb - \rE \by \|_2$, which is a constant factor larger
than the optimal prediction performance
$\|X \bb - \rE \by \|_2$. By optimizing $\lambda$ and $a$,
it is possible to obtain
a similar result as that of Theorem~\ref{thm:struct-constr}.
However, this requires tuning $\lambda$, which is not as convenient as
tuning $s$ in (\ref{eq:struct-constr}).
Note that both results presented here, and those
in  \cite{BuTsWe07-aos} are superior to the more traditional least squares
regression results with $\lambda$ explicitly fixed 
(for example, theoretical results for AIC). 
This is because one can only
obtain the form presented in  Theorem~\ref{thm:struct-constr}
by tuning $\lambda$. Such tuning is important in real applications.

\subsection{Structured greedy algorithm}

We shall introduce a definition before stating our main results.
\begin{definition}
  Given $\cB \subset 2^\cI$, define
  \[
  \rho_0(\cB)=  \max_{B \in \cB} \rho_+(B) , \qquad
  c_0(\cB)=  \max_{B \in \cB} c(B)
  \]
  and
  \[
  c(\bb,\cB)= \min \left\{ \sum_{j=1}^b c(\bar{B}_j)
      :   \Fr(\bb) \subset \bigcup_{j=1}^b \bar{B}_j \quad (\bar{B}_j \in \cB) \right\} .
  \]
\end{definition}
The following theorem shows that if $c(\bb,\cB)$ is small, then one can
use the structured greedy algorithm to
find a coefficient vector $\beta^{(k)}$ that is competitive to $\bb$,
and the coding complexity $c(\beta^{(k)})$ is not much worse than
that of $c(\bb,\cB)$. This implies that if the original
coding complexity $c(\bb)$ can be approximated by block complexity
$c(\bb,\cB)$, then we can approximately solve (\ref{eq:struct-constr}).

\begin{theorem}\label{thm:greedy-conv1}
  Suppose the coding scheme is sub-additive.
  Consider $\bb$ and $\epsilon$ such that
  \[
  \epsilon \in (0, \|\by\|_2^2 - \|X\bb-\by\|_2^2]
  \]
  and
  \[
  s \geq
  \frac{\rho_0(\cB) c(\bb,\cB)}{\gamma \rho_-(s + c(\bb))}
  \ln \frac{\|\by\|_2^2 - \|X\bb-\by\|_2^2}{\epsilon} .
  \]
  Then at the stopping time $k$, we have
  \[
  \hQ(\beta^{(k)}) \leq \hQ(\bb) + \epsilon .
  \]
\end{theorem}

By Theorem~\ref{thm:struct-constr}, the result in
Theorem~\ref{thm:greedy-conv1} implies that
\begin{align*}
&\|X \beta^{(k)} - \rE \by \|_2 \leq
\|X \bb - \rE \by\|_2
+ \sigma \sqrt{2\ln(6/\eta)}
+ 2\sigma (7.4 (s+c_0(\cB)) + 4.7 \ln (6/\eta) +\epsilon/\sigma^2 )^{1/2} , \\
& \|\beta^{(k)} - \bb \|_2^2 \leq
\frac{1}{\rho_-(s+c_0(\cB)+c(\bb)) n}
\left[ 10 \|X \bb- \rE\by\|_2^2 + 37 \sigma^2  (s+c_0(\cB))
+ 29 \sigma^2 \ln (6/\eta) +2.5 \epsilon  \right]
.
\end{align*}

The result shows that in order to approximate a signal $\bb$ up to
accuracy $\epsilon$,
one needs to use coding complexity $O(\ln (1/\epsilon)) c(\bb,\cB)$.
If $\cB$ contains small blocks and their sub-blocks with equal coding length,
and the coding scheme is block coding generated by $\cB$,
then $c(\bb,\cB)=c(\bb)$. In this case
we need $O(s \ln (1/\epsilon))$ to approximate a signal with coding complexity
$s$.

In order to get rid of the $O(\ln (1/\epsilon))$ factor,
backward greedy strategies can be employed, as shown in various recent
work such as \cite{Zhang08-foba}.
For simplicity, we will not analyze such strategies in this paper.
However, in the following, we present an additional convergence
result for structured greedy algorithm that can be applied to
weakly sparse $p$-compressible signals common in practice.
It is shown that the $\ln (1/\epsilon)$ can be removed
for such weakly sparse signals.

\begin{theorem}\label{thm:greedy-conv2}
  Suppose the coding scheme is sub-additive.
  Given a sequence of targets
  $\bb_j$ such that $\hQ(\bb_0) \leq \hQ(\bb_1) \leq \cdots$ and
  $c(\bb_j,\cB) \leq c(\bb_0,\cB)/2^j$.
  If
  \[
  s \geq
  \frac{\rho_0(\cB)}{\gamma \min_j \rho_-(s + c(\bb_j))}
  c(\bb_0,\cB)
  \left[ 3.4 + \sum_{j=0}^\infty 2^{-j}
  \ln \frac{\hQ(\bb_{j+1})-\hQ(\bb_0)+\epsilon}{\hQ(\bb_j)-\hQ(\bb_0)+\epsilon}
  \right]
  \]
  for some $\epsilon>0$.
  Then at the stopping time $k$, we have
  \[
  \hQ(\beta^{(k)}) \leq \hQ(\bb_0) +  \epsilon .
  \]
\end{theorem}

In the above theorem, we can see that if the signal is only weakly sparse, in that $(\hQ(\bb_{j+1})-\hQ(\bb_0)+\epsilon)/(\hQ(\bb_j)-\hQ(\bb_0)+\epsilon)$ grows sub-exponentially in $j$, then we can choose $s= O(c(\bb_0,\cB))$.
This means that
we can find $\beta^{(k)}$ of complexity $s=O(c(\bb_0,\cB))$
to approximate a signal $\bb_0$.
The worst case scenario is when
$\hQ(\bb_{1}) \approx \hQ(0)$, which reduces to the
$s=O(c(\bb_0,\cB) \log(1/\epsilon))$ complexity in Theorem~\ref{thm:greedy-conv1}.

As an application, we introduce the following concept of weakly
sparse compressible target that generalizes the corresponding
concept of compressible signal in standard sparsity from the
compressive sensing literature \cite{Donoho06tit}.
\begin{definition}
The target $\rE \by$ is $(a,q)$-compressible with respect to block $\cB$ if
there exist constants $a,q >0$ such that
for each $s >0$, $\exists \bb(s)$ such that $c(\bb(s),\cB) \leq s$ and
\[
\frac{1}{n} \|X \bb(s) - \rE \by\|_2^2
\leq a s^{-q} .
\]
\end{definition}

\begin{corollary} \label{cor:greedy-compressible}
  Suppose that the target is $(a,q)$-compressible with respect to $\cB$.
  Then with probability $1-\eta$, at the stopping time $k$, we have
  \[
  \hQ(\beta^{(k)}) \leq
  \hQ(\bb(s')) + 2 n a/s'^q
  + 2 \sigma^2 [\ln (2/\eta) +1] ,
  \]
  where
  \[
  s' \leq  \frac{s \; \gamma}{(10+ 3q)\rho_0(\cB) }
  \min_{u \leq s'}\rho_-(s + c(\bb(u))) .
  \]
\end{corollary}
If we assume the underlying coding scheme is block coding generated
by $\cB$, then $\min_{u \leq s'} \rho_-(s + c(\bb(u)))
\leq \rho_-(s + s')$.
The corollary shows that we can approximate a compressible
signal of complexity
$s'$ with complexity $s=O(q s')$ using greedy algorithm.
This means the greedy algorithm obtains optimal rate for weakly-sparse
compressible signals. The sample complexity suffers only a constant factor
$O(q)$.
Combine this result with Theorem~\ref{thm:struct-constr}, and take union bound,
we have with probability $1-2\eta$, at stopping time $k$:
\begin{align*}
&\frac{1}{\sqrt{n}} \|X \beta^{(k)} - \rE \by \|_2 \leq
\sqrt{\frac{a}{s'^q}}
+ \sigma \sqrt{\frac{2\ln(6/\eta)}{n}}
+ 2\sigma \sqrt{\frac{7.4 (s+c_0(\cB)) + 6.7 \ln (6/\eta)}{n} +\frac{2 a}{\sigma^2 s'^q}} , \\
& \|\beta^{(k)} - \bb(s') \|_2^2 \leq
\frac{1}{\rho_-(s+ s' + c_0(\cB)) }
\left[ \frac{15 a}{s'^q} + \frac{37 \sigma^2  (s+c_0(\cB))
+ 34 \sigma^2 \ln (6/\eta)}{n} \right] .
\end{align*}
Given a fixed $n$, we can obtain a convergence result by choosing
$s$ (and thus $s'$) to optimize the right hand side.
The resulting rate is optimal for the special case
of standard sparsity, which
implies that the bound has the optimal form for structured
$q$-compressible targets.
In particular, in compressive sensing applications where $\sigma=0$,
we obtain when sample size reaches $n=O(q s')$, the reconstruction performance
is
\[
\|\bb^{(k)} - \bb\|_2^2 = O(a / s'^q) ,
\]
which matches that of the constrained coding complexity regularization
method in (\ref{eq:struct-constr})
up to a constant $O(q)$.
Since many real data involve weakly sparse signals, our result provides
strong theoretical justification for the use of OMP in such
problems. Our experiments are consistent with the theory.

\section{Experiments}

The purpose of these experiments is to demonstrate the advantage
of structured sparsity over standard sparsity. We compare the
proposed StructOMP to OMP and Lasso, which are standard algorithms
to achieve sparsity but without considering structure. In our
experiments, we use Lasso-modified least angle regression
(LAS/Lasso) as the solver of Lasso \cite{Efron04}. 
In order to
quantitatively compare performance of different algorithms, we use
recovery error, defined as the relative difference in 2-norm
between the estimated sparse coefficient vector
$\hat{\beta}_{est}$ and the ground-truth sparse coefficient $\bb$:
$\|\hat{\beta}_{est}-\bb\|_{2}/\|\bb\|_{2}$. Our experiments focus
on graph sparsity, with several different underlying graph
structures. Note that graph sparsity is more general than group
sparsity; in fact connected regions may be regarded as dynamic
groups that are not pre-defined. However, for illustration,
we include a comparison with group Lasso using some 1D simulated 
examples, where the underlying structure can be more easily approximated
by pre-defined groups. 
Since additional experiments involving more complicated structures are more difficult to approximate by pre-defined groups, we exclude group-Lasso in those
experiments.

\subsection{Simulated 1D Signals with Line-Structured Sparsity}

In the first experiment, we randomly generate a $1D$ structured
sparse signal with values $\pm 1$, where $p=512$, $k=64$ and
$g=4$. The support set of these signals is composed of \emph{g}
connected regions. Here, each component of the sparse coefficient
is connected to two of its adjacent components, which forms the
underlying graph structure. The graph sparsity concept introduced
earlier is used to compute the coding length of sparsity patterns
in StructOMP. The projection matrix $X$ is generated by creating
an $n \times p$ matrix with i.i.d. draws from a standard Gaussian
distribution $N(0, 1)$. For simplicity, the rows of $X$ are
normalized to unit magnitude. Zero-mean Gaussian noise with
standard deviation $\sigma=0.01$ is added to the measurements. Our
task is to compare the recovery performance of StructOMP to those
of OMP, Lasso and group Lasso for these structured sparsity
signals.

\begin{figure}[htbp]
\centering \vspace{-0.2cm}
\includegraphics[width = 1.0 \columnwidth]{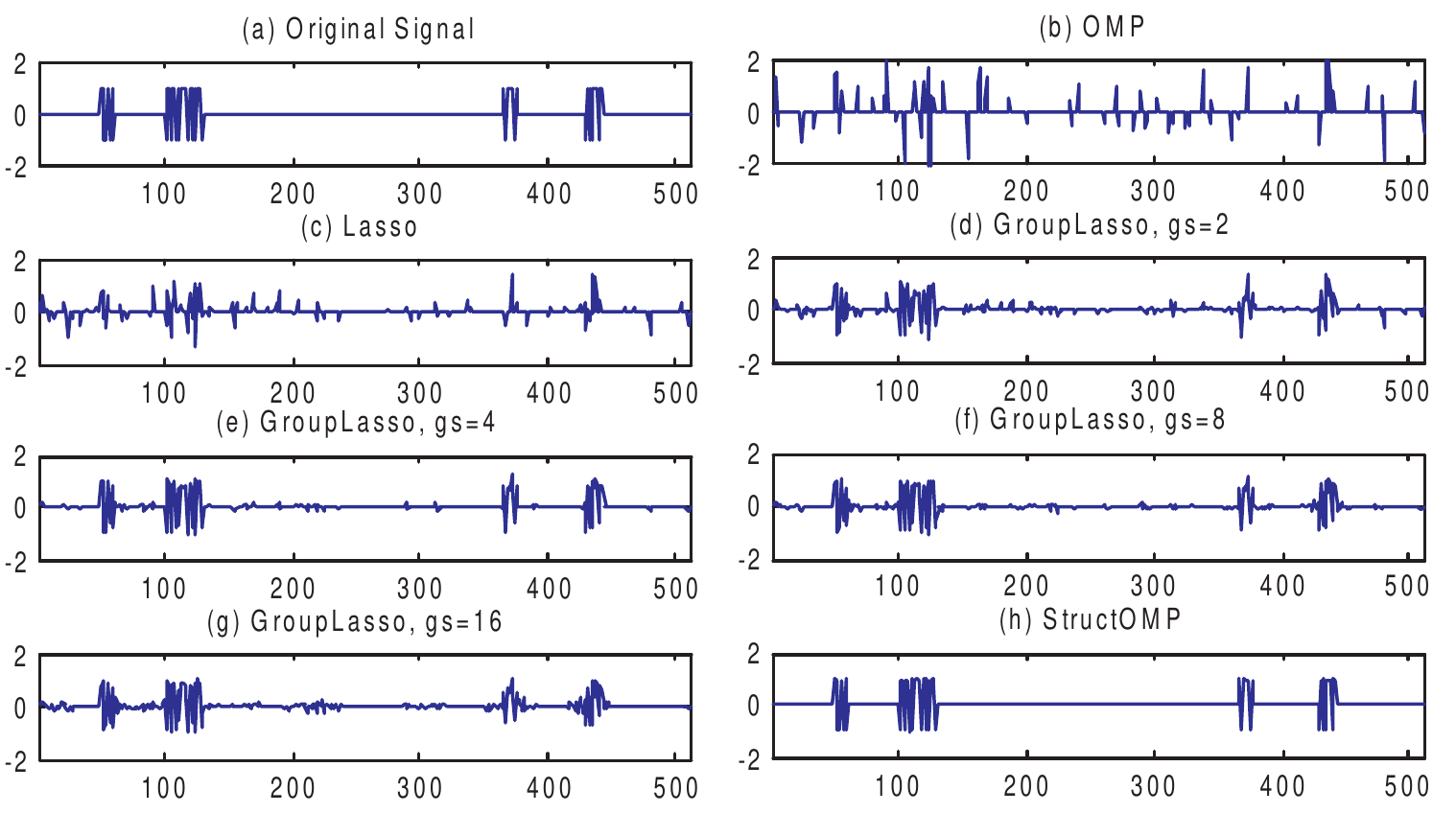}
\caption{Recovery results of 1D signal with graph-structured
sparsity. (a) original data; (b) recovered results with OMP (error
is 0.9921); (c) recovered results with Lasso (error is 0.8660);;
(d) recovered results with Group Lasso (error is 0.4832 with group
size gs=2); (e) recovered results with Group Lasso (error is
0.4832 with group size gs=4);(f) recovered results with Group
Lasso (error is 0.2646 with group size gs=8);(g) recovered results
with Group Lasso (error is 0.3980 with group size gs=16); (h)
recovered results with StructOMP (error is 0.0246).}
\label{fig:Exp1D_example} \vspace{-0.2cm}
\end{figure}

\begin{figure}[htbp]
\centering \vspace{-0.0cm}
    \subfigure[]{\label{fig:Exp1D_M_stat_rate}
        \includegraphics[scale=0.52]{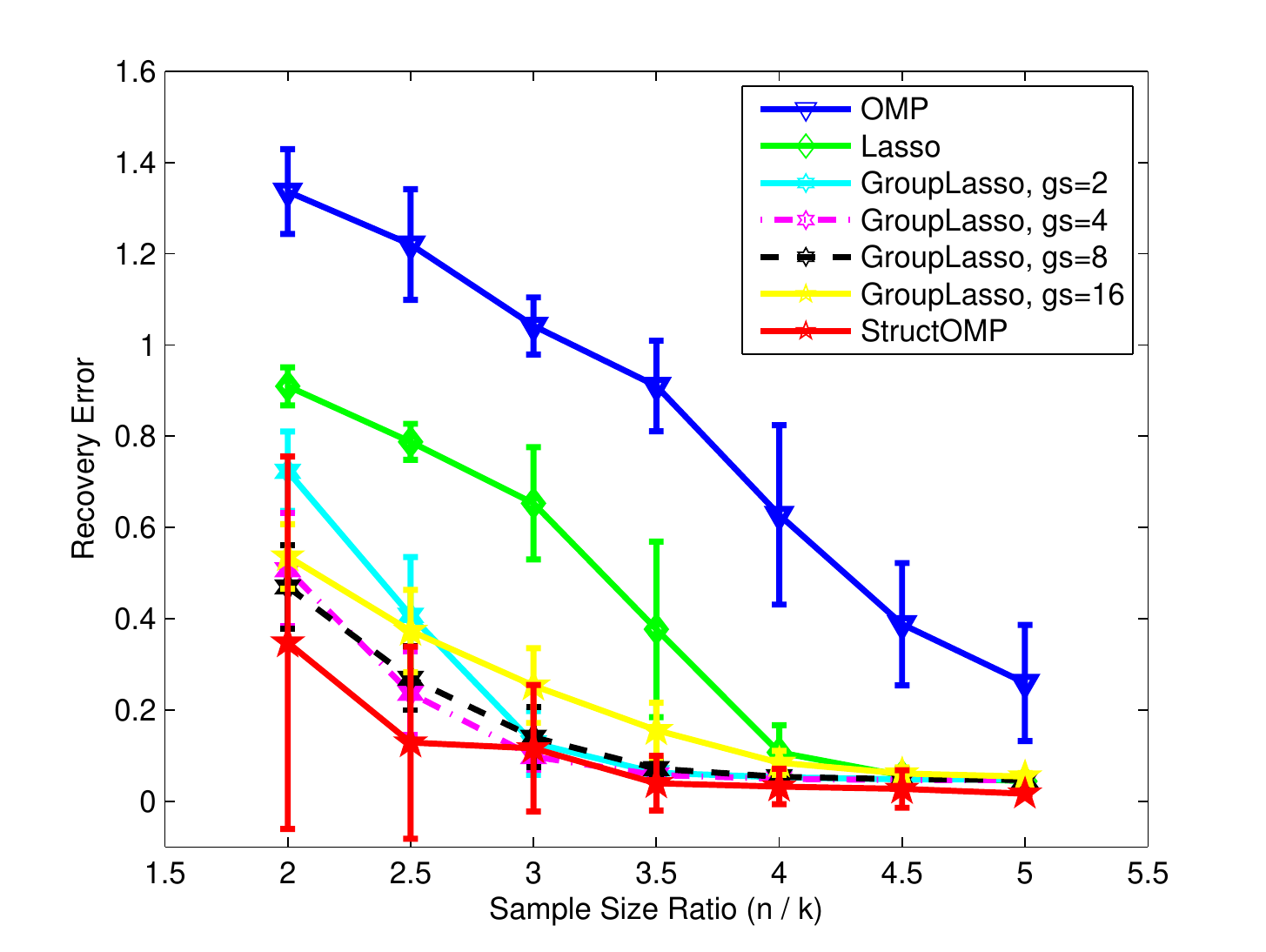}}
    \subfigure[]{\label{fig:Exp1D_M_stat_Weak_rate}
        \includegraphics[scale=0.52]{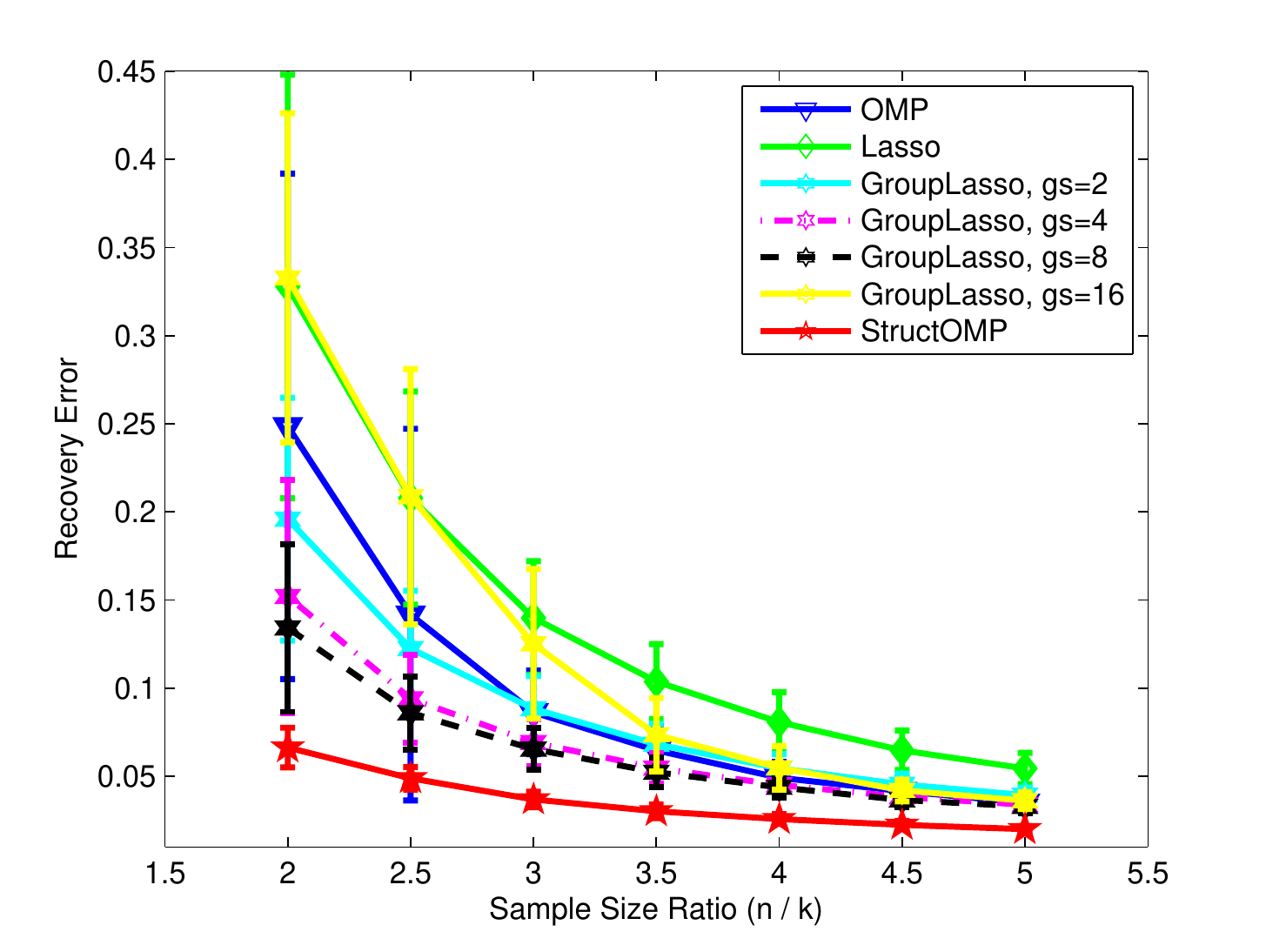}}
\caption{Recovery error vs. Sample size ratio $(n/k)$: a) 1D
signals; (b) 1D Weak sparse
signal}\label{fig:Exp1D_M_stat}\vspace{-0.0cm}
\end{figure}

Figure~\ref{fig:Exp1D_example} shows one instance of generated
signal and the corresponding recovered results by different
algorithms when $n=160$. Since the sample size $n$ is not big
enough, OMP and Lasso do not achieve good recovery results,
whereas the StructOMP algorithm achieves near perfect recovery of
the original signal. We also include group Lasso in this experiment
for illustration. We use pre-defined consecutive groups that
do not completely overlap with the support of the signal.
Since we do not know the correct group size, we just try group Lasso 
with several different group
sizes (gs=2, 4, 8, 16). Although the results obtained with group
Lasso are better than those of OMP and Lasso, they are still
inferior to the results with StructOMP. As mentioned, this is because
the pre-defined groups do not completely overlap with the support of
the signal, which reduces the efficiency.
To study how the sample
size $n$ affects the recovery performance, we vary the sample size
and record the recovery results by different algorithms. To reduce
the randomness, we perform the experiment 100 times for each
sample size. Figure~\ref{fig:Exp1D_M_stat_rate} shows the recovery
performance of the three algorithms, averaged over 100 random runs
for each sample size. As expected, StructOMP is better than the
group Lasso and far better than the OMP and Lasso. The results
show that the proposed StructOMP can achieve better recovery
performance for structured sparsity signals with less samples.

\begin{figure}[htbp]
\centering \vspace{-0.2cm}
\includegraphics[width = 1 \columnwidth]{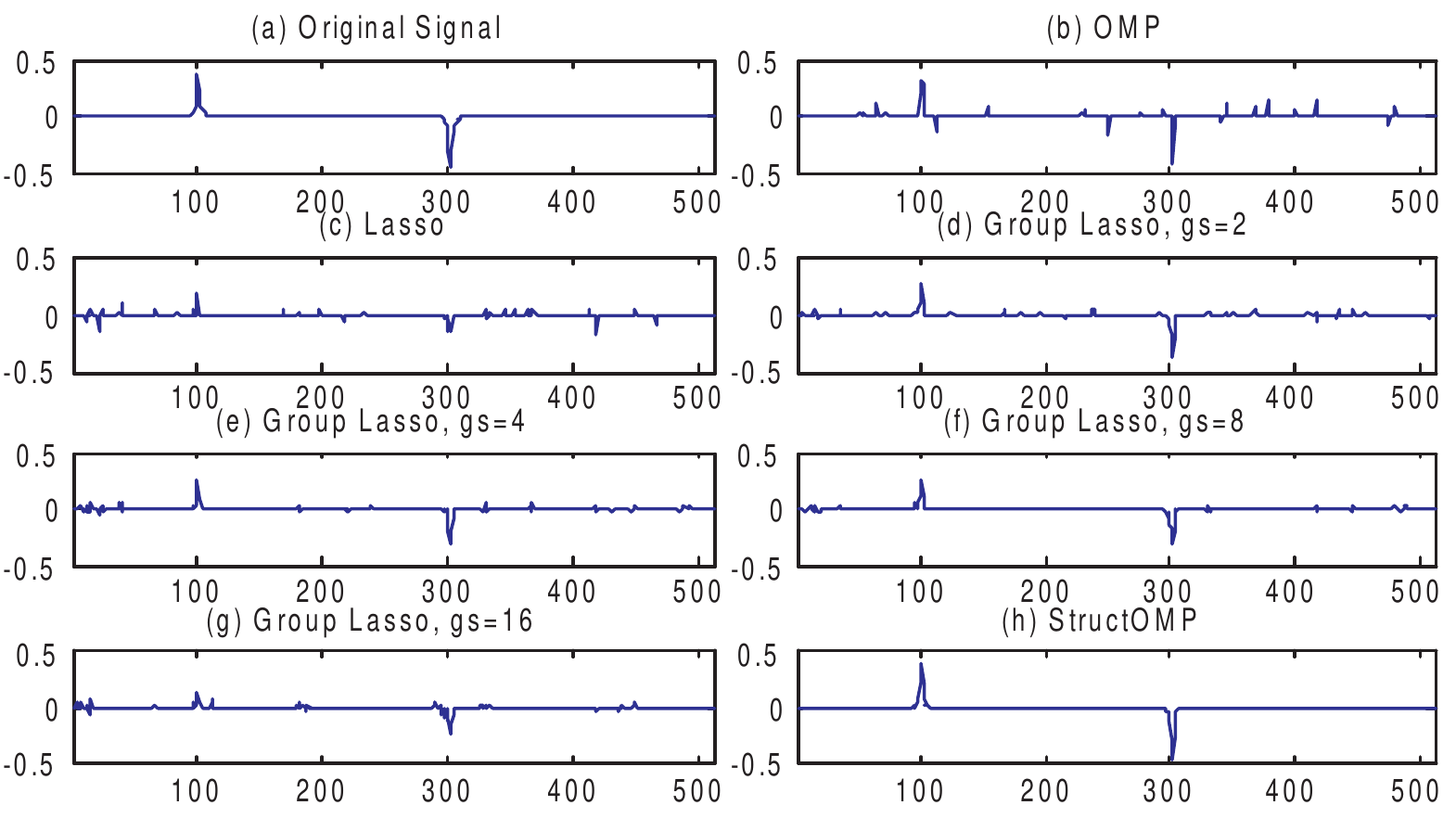}
\caption{Recovery results of 1D weakly sparse signal with
line-structured sparsity. (a) original data; (b) recovered results
with OMP (error is 0.5599); (c) recovered results with Lasso
(error is 0.6686);; (d) recovered results with Group Lasso (error
is 0.4732 with group size gs=2); (e) recovered results with Group
Lasso (error is 0.2893 with group size gs=4);(f) recovered results
with Group Lasso (error is 0.2646 with group size gs=8);(g)
recovered results with Group Lasso (error is 0.5459 with group
size gs=16); (h) recovered results with StructOMP (error is
0.0846).} \label{fig:Exp1DWeak_example} \vspace{-0.2cm}
\end{figure}

Note that Lasso performs better than OMP in the first example.
This is because the signal is strongly sparse (that is, all
nonzero coefficients are significantly different from zero). 
In
the second experiment, we randomly generate a $1D$ structured
sparse signal with weak sparsity, where the nonzero coefficients
decay gradually to zero, but there is no clear cutoff. 
As expected in our theory (see Theorem~\ref{thm:greedy-conv2} and
discussions thereafter), OMP becomes much more competitive
relative to Lasso. In fact OMP has better performance than Lasso
in this more realistic situation.
One instance of generated signal is shown in Figure
\ref{fig:Exp1DWeak_example} (a). Here, $p=512$ and all coefficient
of the signal are not zeros. We define the sparsity $k$ as the
number of coefficients that contain $95\%$ of the image energy.
The support set of these signals is composed of $g=2$ connected
regions. Again, each element of the sparse coefficient vector is
connected to two of its adjacent elements, which forms the
underlying 1D line graph structure. The graph sparsity concept
introduced earlier is used to compute the coding length of
sparsity patterns in StructOMP. The projection matrix $X$ is
generated by creating an $n \times p$ matrix with i.i.d. draws
from a standard Gaussian distribution $N(0, 1)$. For simplicity,
the rows of $X$ are normalized to unit magnitude. Zero-mean
Gaussian noise with standard deviation $\sigma=0.01$ is added to
the measurements. 

Figure~\ref{fig:Exp1DWeak_example} shows one generated signal and
its recovered results by different algorithms when $k=32$ and
$n=48$. Again, we observe that OMP and Lasso do not achieve good
recovery results, whereas the StructOMP algorithm achieves near
perfect recovery of the original signal. 
We try group Lasso with
several different group sizes (gs=2, 4, 8, 16). Although the
results obtained with group Lasso are better than those of OMP and
Lasso, they are still inferior to the results with StructOMP. In
order to study how the sample size $n$ effects the recovery
performance, we vary the sample size and record the recovery
results by different algorithms. To reduce the randomness, we
perform the experiment 100 times for each of the sample sizes.
Figure~\ref{fig:Exp1D_M_stat_Weak_rate} shows the recovery
performance of different algorithms, averaged over 100 random runs
for each sample size. As expected, StructOMP algorithm is superior
in all cases. What's different from the first experiment is that
the recovery error of OMP becomes smaller than that of Lasso. This
result is consistent with our theory, which predicts that if the
underlying signal is weakly sparse, then the relatively
performance of OMP becomes comparable to Lasso.

\subsection{2D Image Compressive Sensing with Tree-structured Sparsity}

\begin{figure}[htbp]
\centering \vspace{-0.0cm}
    \subfigure[]{\label{fig:Exp2D_Tree_example1}
        \includegraphics[scale=0.23]{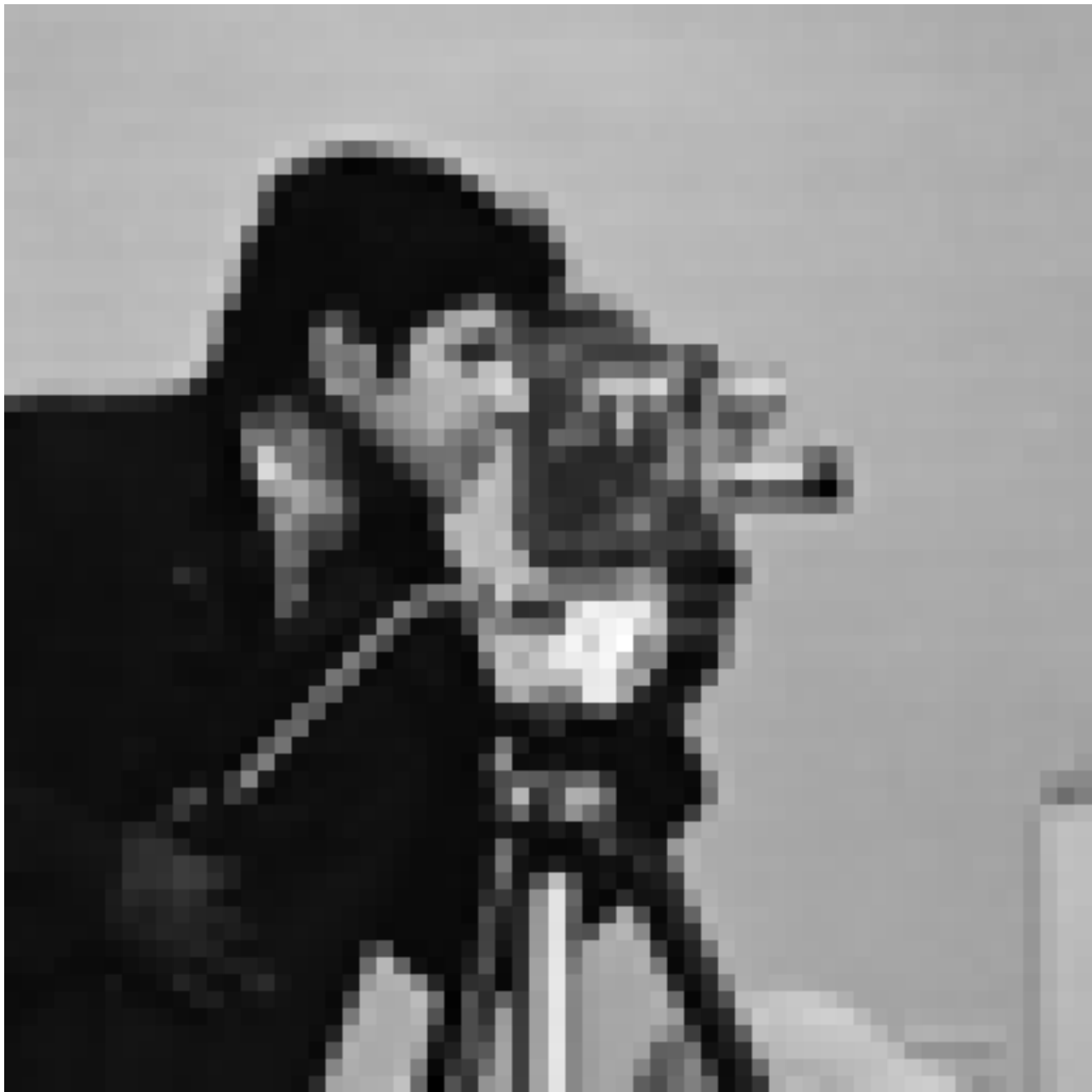}}
    \subfigure[]{\label{fig:Exp2D_Tree_example2}
        \includegraphics[scale=0.23]{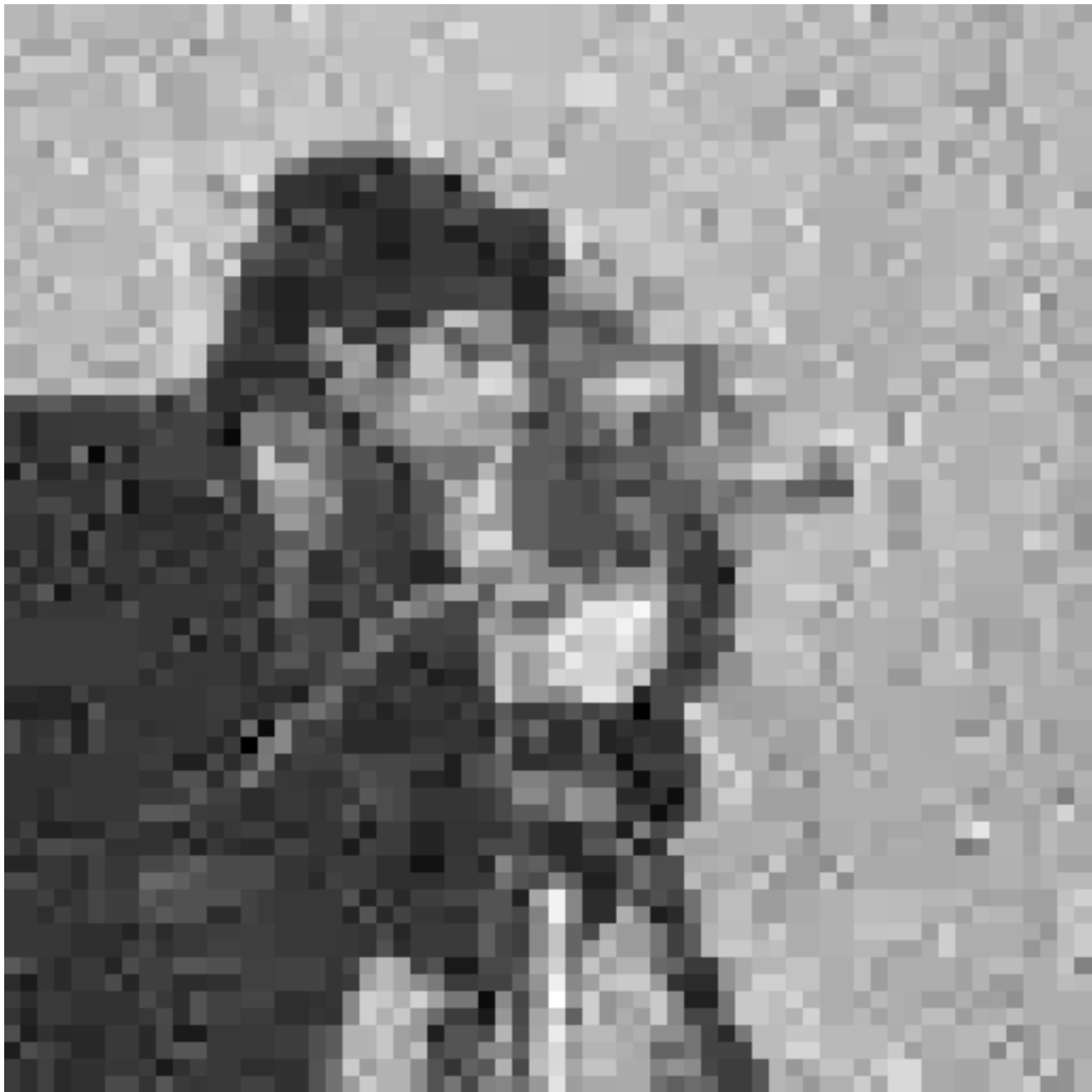}}
    \subfigure[]{\label{fig:Exp2D_Tree_example3}
        \includegraphics[scale=0.23]{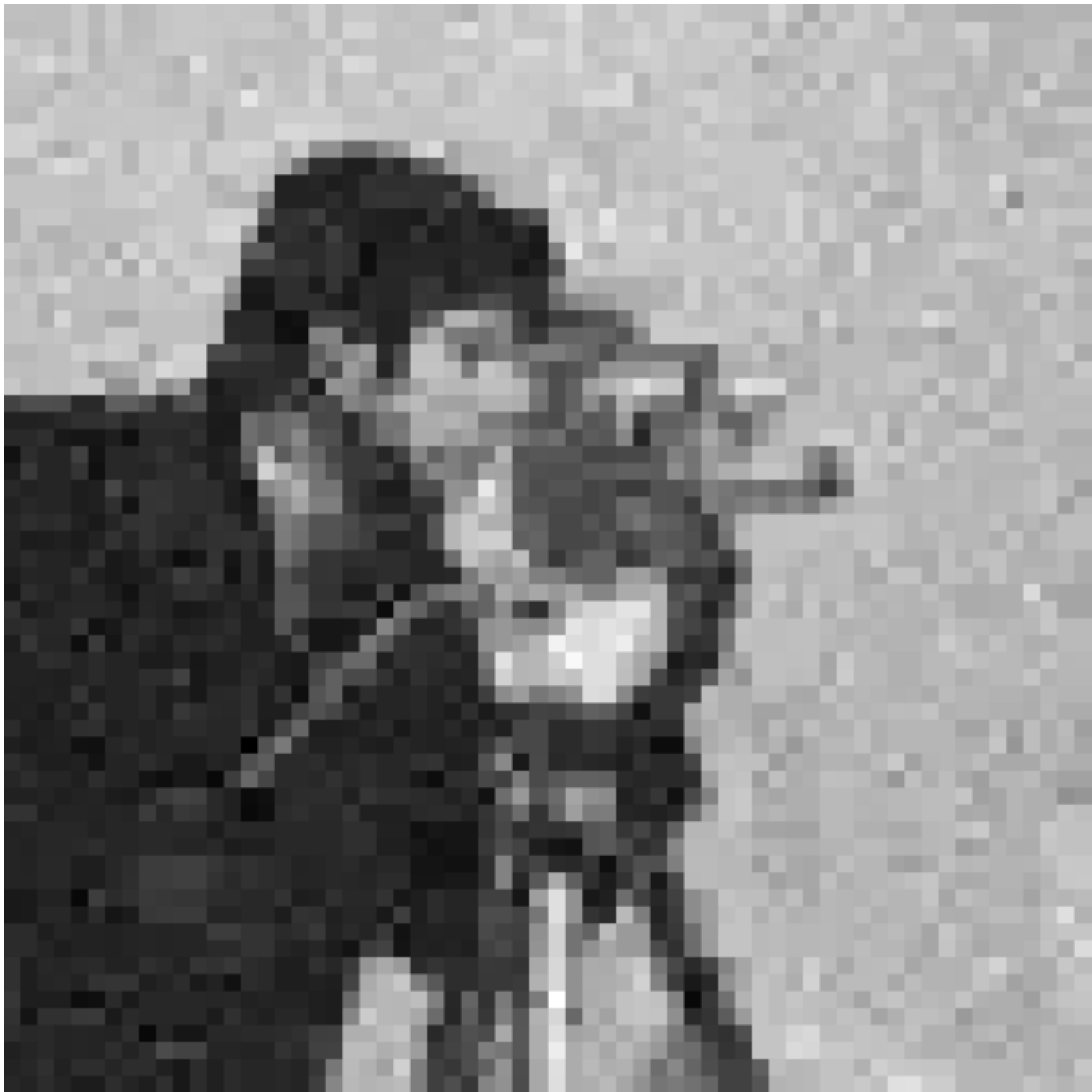}}
    \subfigure[]{\label{fig:Exp2D_Tree_example4}
        \includegraphics[scale=0.23]{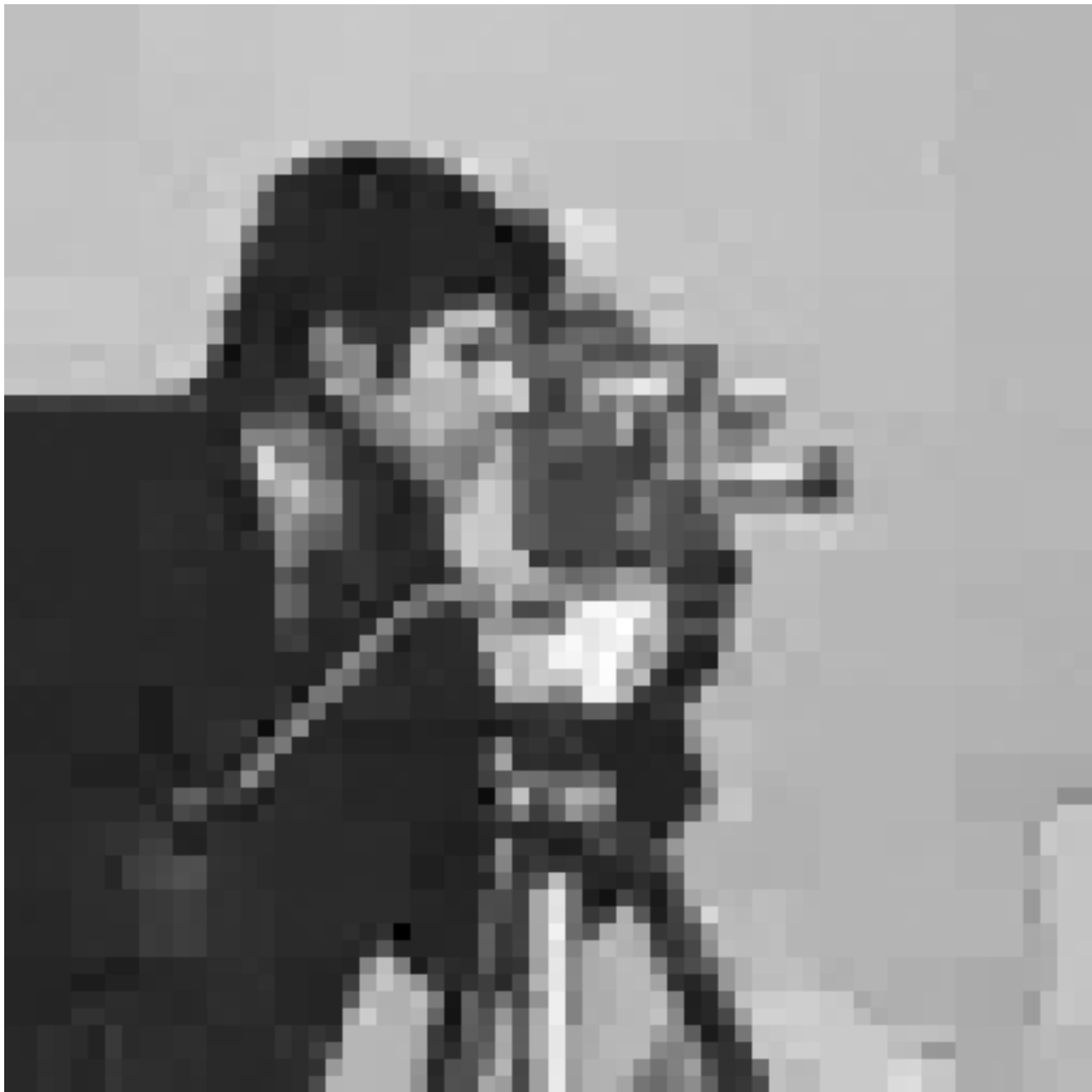}}
\caption{Recovery results with sample size $n=2048$: (a) the
background subtracted image, (b) recovered image with OMP (error
is 0.21986), (c) recovered image with Lasso (error is 0.1670) and
(d) recovered image with StructOMP (error is
0.0375)}\label{fig:Exp2D_Tree_example}\vspace{-0.0cm}
\end{figure}

It is well known that 2D natural images are sparse in a
wavelet basis. Their wavelet coefficients have a hierarchical tree
structure, which is widely used for wavelet-based
compression algorithms \cite{Shapiro93}.
Figure \ref{fig:Exp2D_Tree_example1} shows
a widely used example image with size $64 \times 64$:
\emph{cameraman}. Each 2D wavelet coefficient of this image is
connected to its parent coefficient and child coefficients, which
forms the underlying hierarchical tree structure (which is a
special case of graph sparsity). In our experiment, we choose
Haar-wavelet to obtain its tree-structured sparsity wavelet
coefficients. The projection matrix $X$ and noises are generated
with the same method as that for 1D structured sparsity signals.
OMP, Lasso and StructOMP are used to recover the wavelet
coefficients from the random projection samples respectively.
Then, the inverse wavelet transform is used to reconstruct the
images with these recovered wavelet coefficients. Our task is to
compare the recovery performance of the StructOMP to those of OMP
and Lasso.

\begin{figure}[htbp]
\centering \vspace{-0.0cm}
    \subfigure[]{\label{fig:Exp2D_M_stat_Tree_rate}
        \includegraphics[scale=0.52]{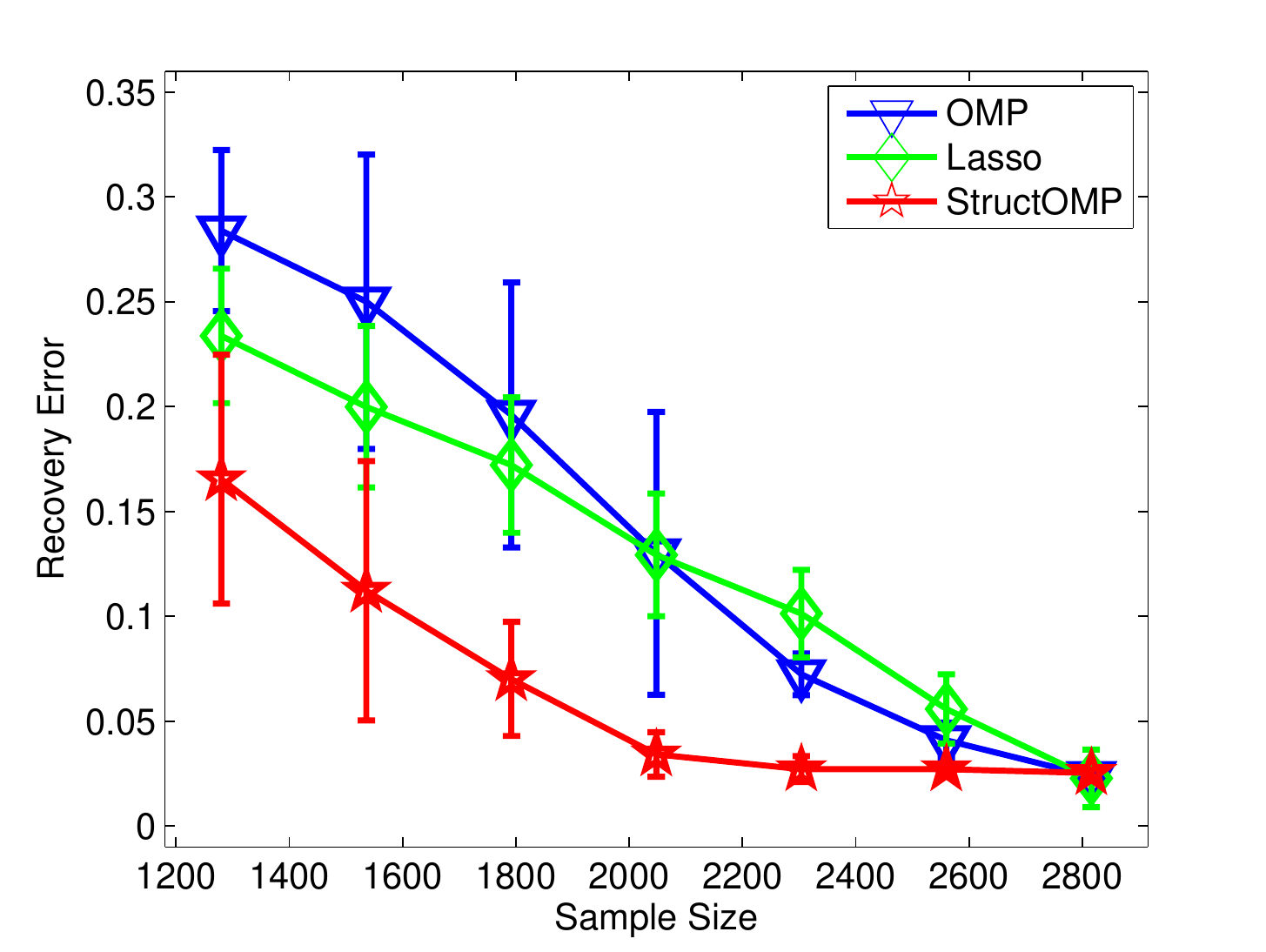}}
    \subfigure[]{\label{fig:Exp2D_M_stat_BG_rate}
        \includegraphics[scale=0.52]{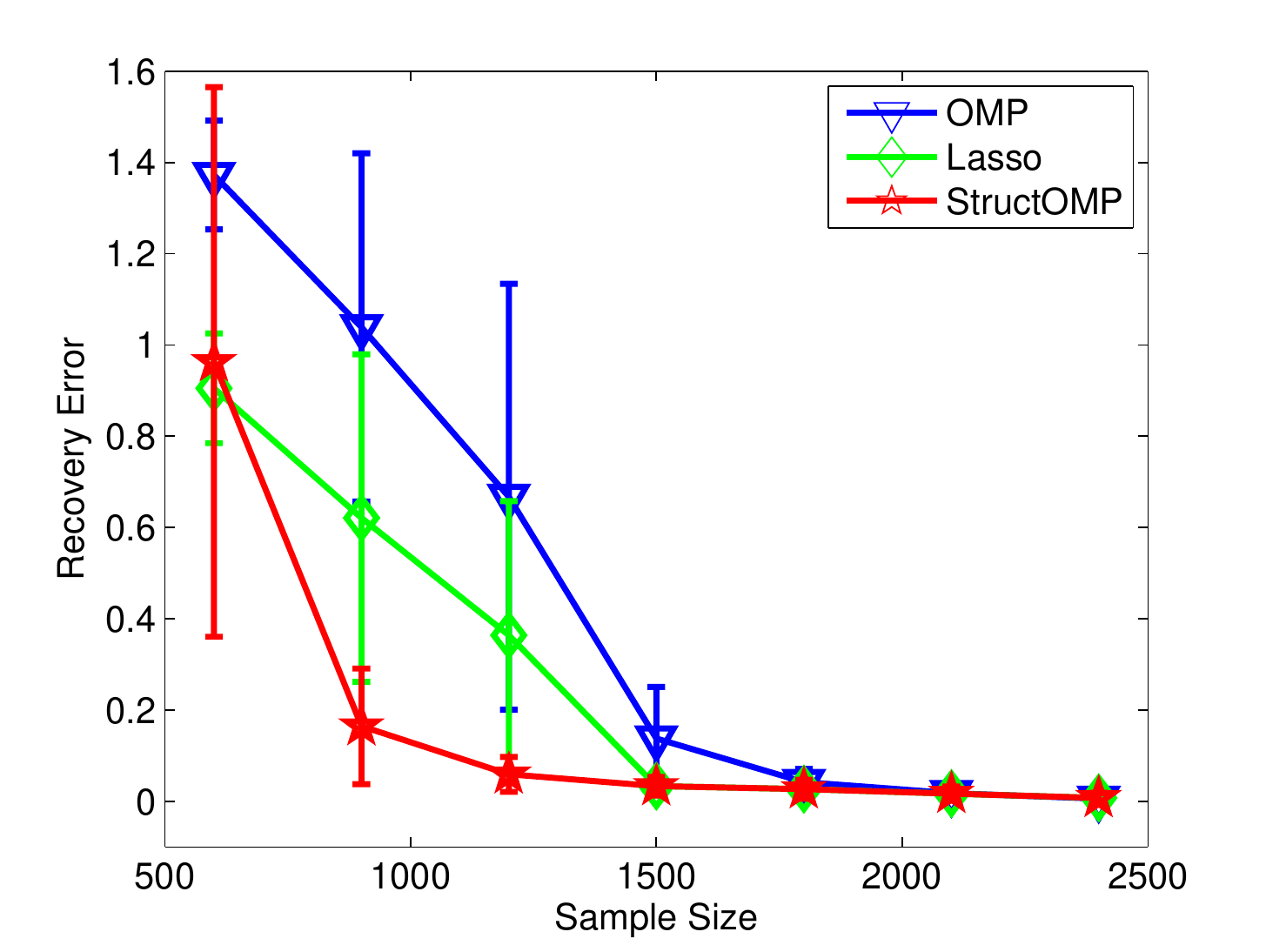}}
\caption{Recovery error vs. Sample size: a) 2D image with
tree-structured sparsity in wavelet basis; (b) background
subtracted images with structured
sparsity}\label{fig:Exp2D_BG_stat}\vspace{-0.0cm}
\end{figure}

Figure \ref{fig:Exp2D_Tree_example} shows one example of the
recovered results by different algorithms. It shows that StructOMP
obtains the best recovered result.
Figure~\ref{fig:Exp2D_M_stat_Tree_rate} shows the recovery
performance of the three algorithms, averaged over 100 random runs
for each sample size. The StructOMP algorithm is better than both
Lasso and OMP in this case.
Since real image data are weakly sparse,
the performance of standard OMP (without structured
sparsity) is similar to that of Lasso.

\vspace{-0.0cm}
\subsection{Background Subtracted Images for Robust Surveillance}

\begin{figure}[htbp]
\centering \vspace{-0.0cm}
    \subfigure[]{\label{fig:Exp2D_BG_example1}
        \includegraphics[scale=0.26]{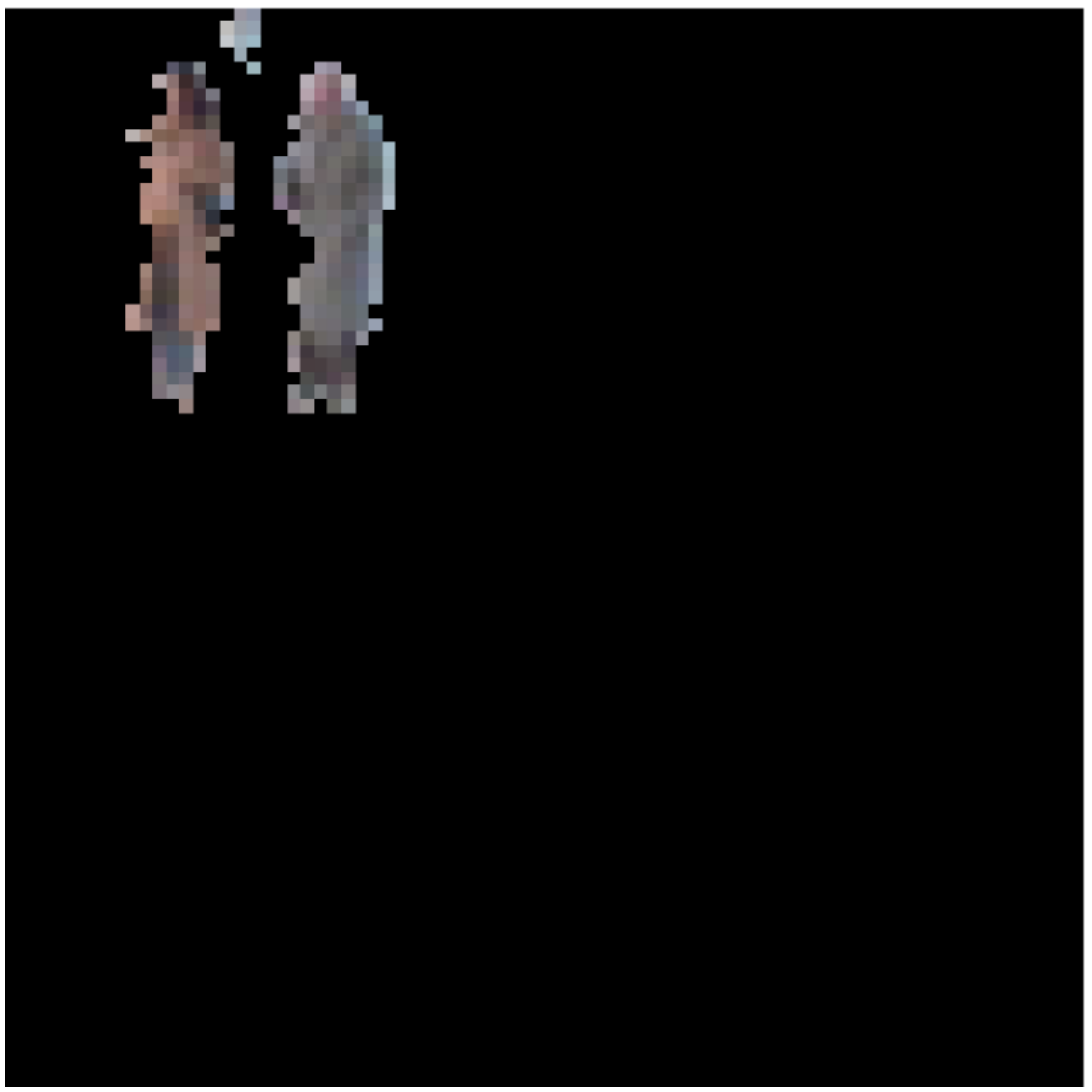}}
    \subfigure[]{\label{fig:Exp2D_BG_example2}
        \includegraphics[scale=0.26]{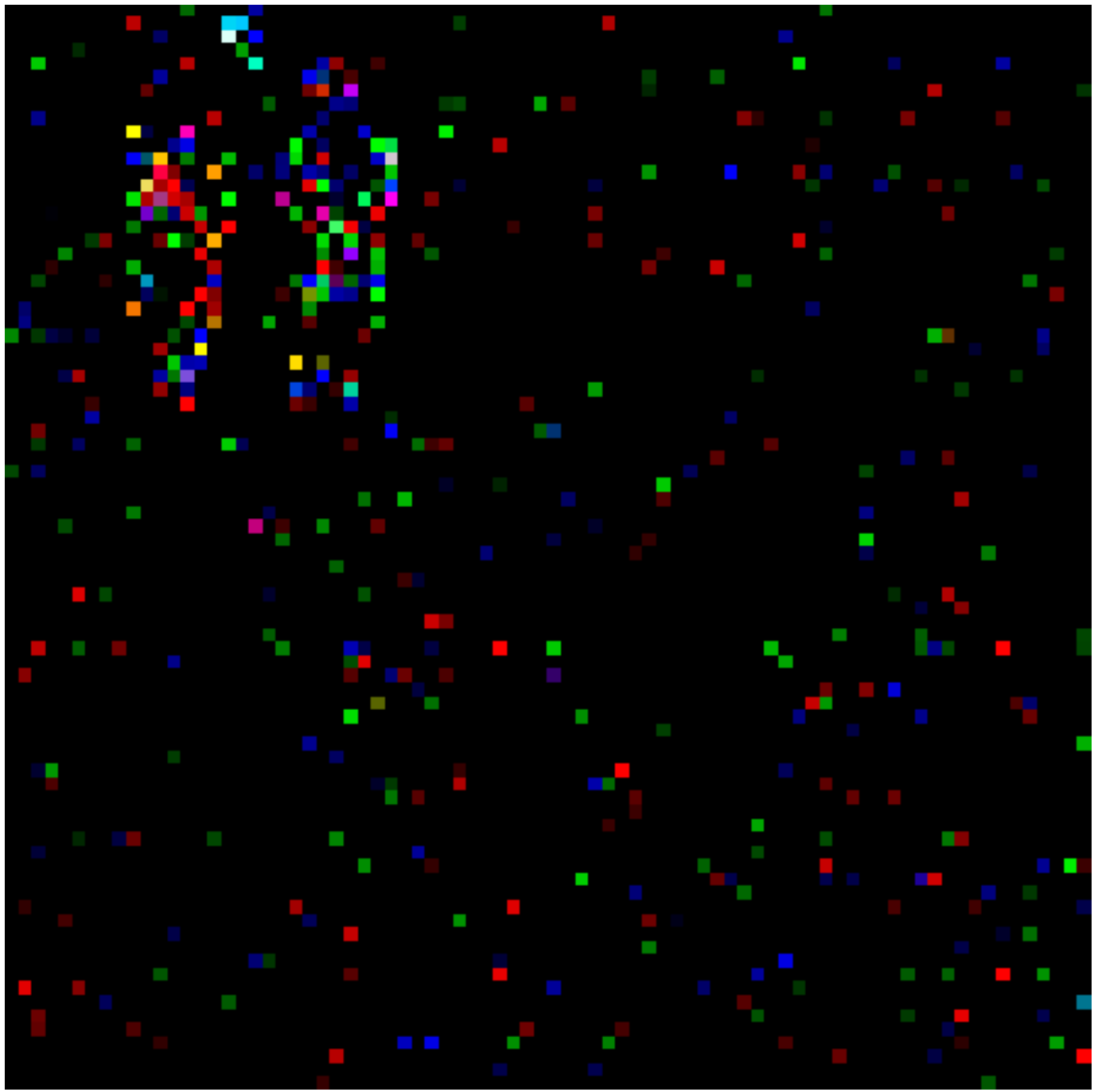}}
    \subfigure[]{\label{fig:Exp2D_BG_example3}
        \includegraphics[scale=0.26]{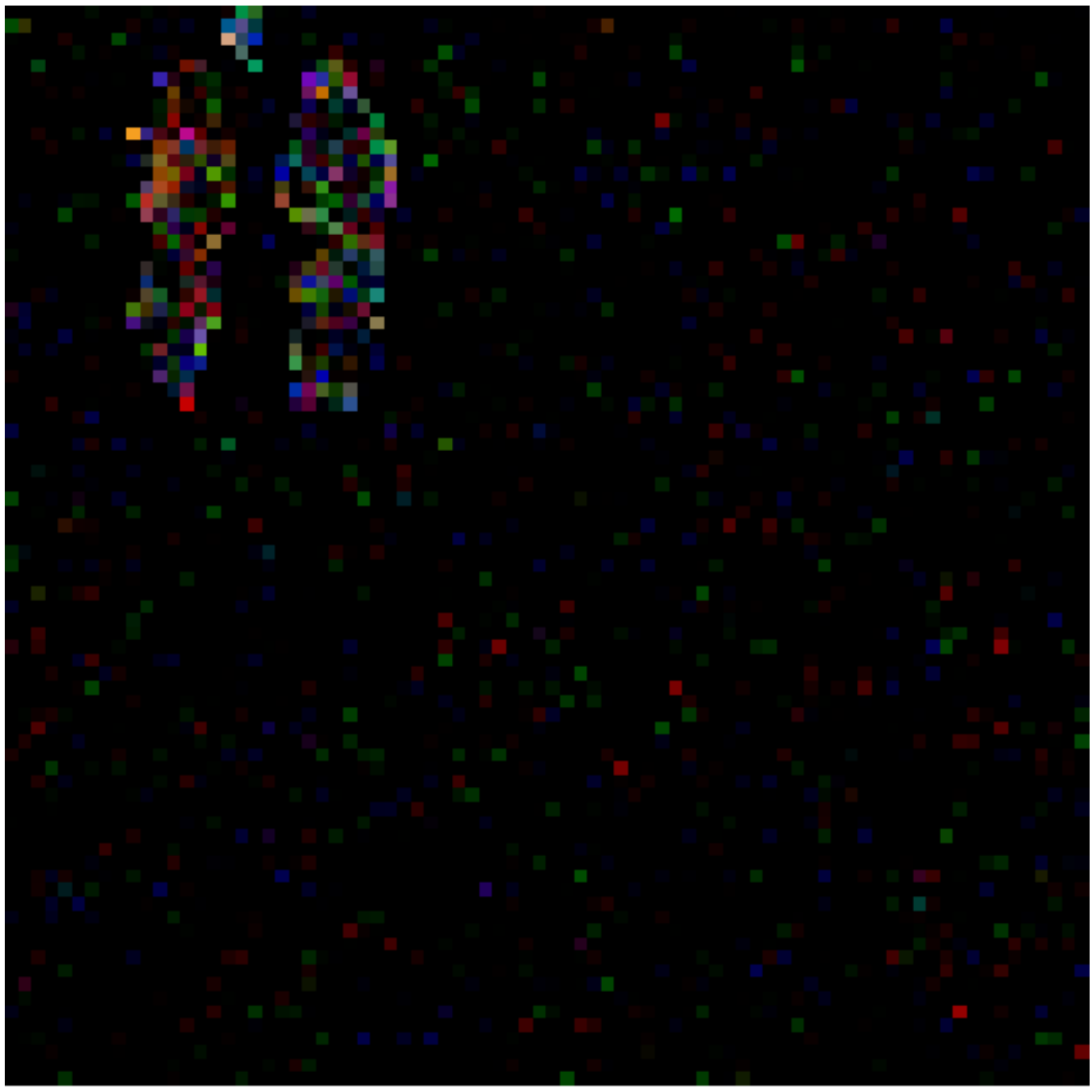}}
    \subfigure[]{\label{fig:Exp2D_BG_example4}
        \includegraphics[scale=0.26]{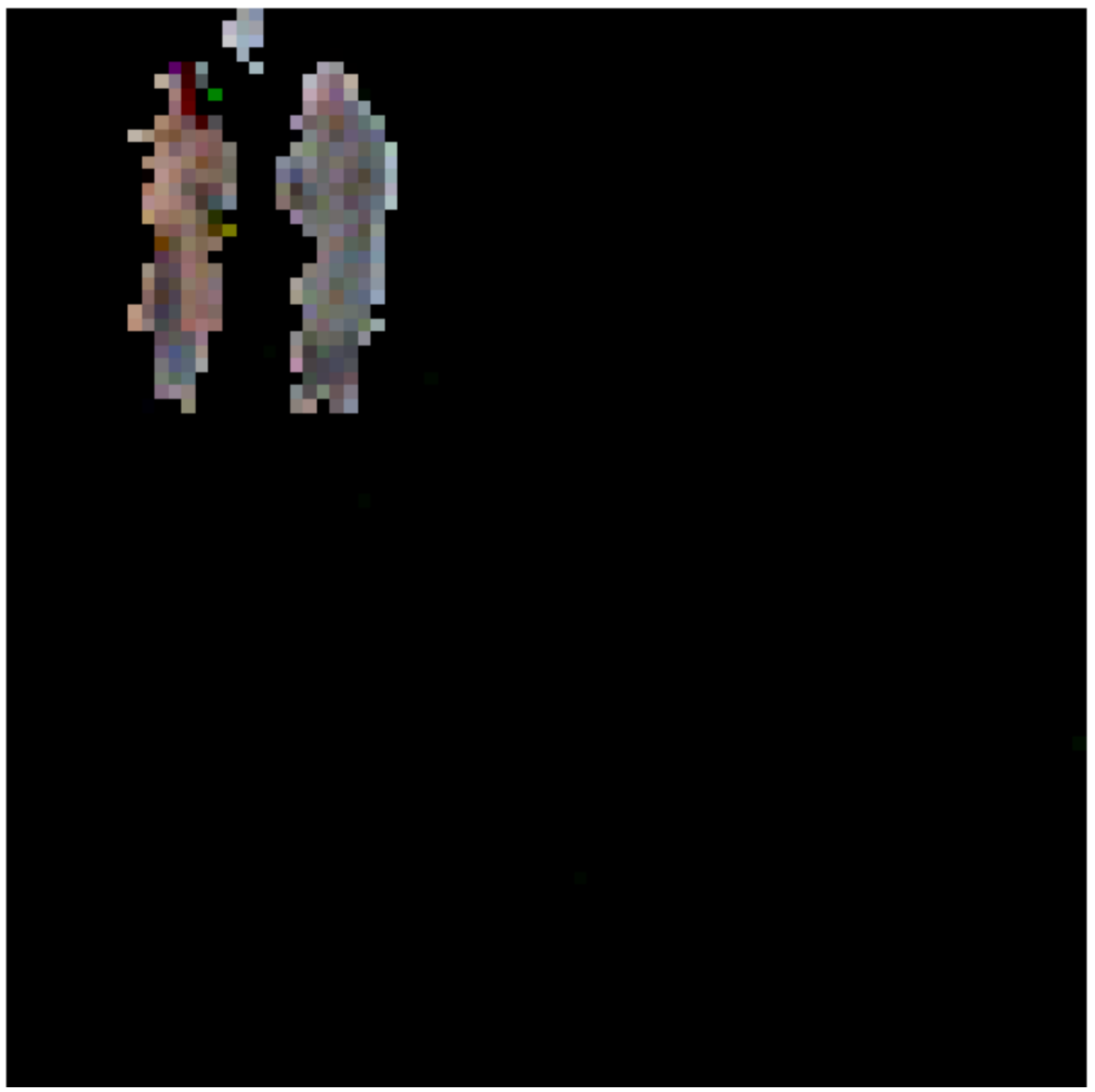}}
\caption{Recovery results with sample size $n=900$: (a) the
background subtracted image, (b) recovered image with OMP (error
is 1.1833), (c) recovered image with Lasso (error is 0.7075) and
(d) recovered image with StructOMP (error is
0.1203)}\label{fig:Exp2D_BG_example}\vspace{-0.0cm}
\end{figure}

Background subtracted images are typical structure sparsity data
in static video surveillance applications. They generally
correspond to the foreground objects of interest. Unlike the whole
scene, these images are not only spatially sparse but also
inclined to cluster into groups, which correspond to different
foreground objects. Thus, the StructOMP algorithm can obtain superior
recovery from compressive sensing measurements that are received
by a centralized server from multiple and randomly placed
optical sensors. In this experiment,
the testing video is downloaded from
http://homepages.inf.ed.ac.uk/rbf/CAVIARDATA1/. The background
subtracted images are obtained with the software
\cite{Zivkovic06pr}. One sample image frame is shown in Figure
\ref{fig:Exp2D_BG_example1}. The support set of 2D images is thus
composed of several connected regions. Here, each pixel of the 2D
background subtracted image is connected to four of its adjacent
pixels, forming the underlying graph structure in graph sparsity.
The results shown in Figure \ref{fig:Exp2D_BG_example} demonstrate
that the StructOMP outperforms both OMP and Lasso in recovery. We
randomly choose 100 background subtracted images as test images.
Figure \ref{fig:Exp2D_M_stat_BG_rate} shows the recovery
performance as a function of increasing sample sizes. It
demonstrates again that StructOMP significantly outperforms OMP
and Lasso in recovery performance on video data.
Comparing to the image compression example in the previous section,
the background subtracted
images have a more clearly defined sparsity pattern where nonzero coefficients are generally distinct from zero (that is, stronger sparsity); this explains why Lasso performs better than the standard (unstructured) OMP on this particular data. The result is again consistent with our theory.

\section{Discussion}

This paper develops a theory for structured sparsity where prior knowledge
allows us to prefer certain sparsity patterns to others.
Some examples are presented to illustrate the concept.
The general framework established in this paper includes the recently
popularized group sparsity idea has a special case.

In structured sparsity, the complexity of learning is measured by
the coding complexity $c(\bb) \leq \|\bb\|_0 + \cl(\Fr(\bb))$ instead
of $\|\bb\|_0 \ln p$ which determines the complexity in standard sparsity.
Using this notation, a theory parallel to that of
the standard sparsity is developed.
The theory shows that
if the coding length $\cl(\Fr(\bb))$ is small for a target coefficient
vector $\bb$, then the complexity of learning $\bb$ can be
significantly smaller than the corresponding complexity in standard
sparsity. Experimental results demonstrate that significant improvements
can be obtained on some real problems that have natural structures.

The structured greedy algorithm presented in this paper is the first efficient
algorithm proposed to handle the general structured sparsity learning.
It is shown that the algorithm is effective under appropriate conditions.
Future work include additional computationally efficient methods such as
convex relaxation methods (e.g. $L_1$ regularization for standard sparsity,
and group Lasso for strong group sparsity) and backward greedy strategies
to improve the forward greedy method considered in this paper.

\bibliographystyle{plain}
\bibliography{group}

\appendix
\section{Proof of Proposition~\ref{prop:noise}}

\begin{lemma}  \label{lem:sub-Gaussian}
  Consider a fixed vector $\bx \in \R^n$, and a random vector
  $\by \in \R^n$ with independent sub-Gaussian components:
  $\rE e^{t (\by_i -\rE \by_i)} \leq e^{\sigma^2 t^2/2}$ for all $t$ and $i$,
  then $\forall \epsilon>0$:
  \[
  \pr \left( \left|\bx^\top \by - \rE \bx^\top \by \right| \geq \epsilon\right)
  \leq 2 e^{-\epsilon^2/(2\sigma^2 \|\bx\|_2^2)} .
  \]
\end{lemma}
\begin{proof}
  Let $s_n=\sum_{i=1}^n (\bx_i \by_i -\rE \bx_i \by_i)$; then by assumption,
  $\rE (e^{t s_n}+e^{-t s_n}) \leq 2 e^{\sum_i \bx_i^2 \sigma^2 t^2/2}$, which implies
 that
  $\pr ( |s_n| \geq \epsilon) e^{ t \epsilon} \leq 2 e^{\sum_i \bx_i^2 \sigma^2 t^2/2
}$.
  Now let $t=\epsilon/(\sum_i \bx_i^2 \sigma^2)$, we obtain the desired bound.
\end{proof}

The following lemma is taken from \cite{Pisier89}. Since the proof is simple,
it is included for completeness.
\begin{lemma}
  Consider the unit sphere $S^{k-1}=\{x: \|x\|_2 = 1\}$ in
  $\mathbb{R}^{k}$ ($k \geq 1$).
  Given any $\varepsilon > 0$,
  there exists an $\varepsilon$-cover  $Q \subset S^{k-1}$
  such that
  $\min_{q \in Q}\|x-q\|_{2}\leq \varepsilon$
  for all  $\|x\|_2 = 1$, with
  $|Q| \leq (1+2/\varepsilon)^{k}$.
  \label{lem:cover}
\end{lemma}
\begin{proof}
Let $B^{k}=\{x: \|x\|_2 \leq 1\}$ be the unit ball in $\mathbb{R}^{k}$.
Let $Q = \{q_{i}\}_{i=1,\ldots, |Q|} \subset S^{k-1}$ be a maximal subset such that
$\|q_{i}-q_{j}\|_{2}>\varepsilon$ for all $i \neq j$.
By maximality, $Q$ is an $\varepsilon$-cover of $S^{k-1}$.
Since the balls
$q_{i}+(\varepsilon/2)B^{k}$ are disjoint and belong to
$(1+\varepsilon/2)B^{k}$, we have
\[
\sum_{i\leq |Q|} vol(q_{i}+(\varepsilon/2)B^{k})\leq
vol((1+\varepsilon/2)B^{k}) .
\]
Therefore,
\[
|Q|(\varepsilon/2)^{k}vol(B^{k}) \leq
(1+\varepsilon/2)^{k}vol(B^{k}),
\]
which implies that $|Q| \leq (1+2/\varepsilon)^{k}$.
\end{proof}

\subsection*{Proof of Proposition~\ref{prop:noise}}

According to Lemma \ref{lem:cover}, given $\epsilon_1>0$,
there exists a finite set $Q=\{q_i\}$ with
$|Q|\leq (1+2/\epsilon_{1})^{k}$
such that $\|Pq_i\|_2=1$ for all $i$, and
$\min_{i}\|P\beta-P q_i\|_{2}\leq \epsilon_1$
for all $\|P\beta\|_2=1$.

For each $i$, Lemma~\ref{lem:sub-Gaussian} implies that
$\forall \epsilon_2>0$:
\[
\pr \left( \left|q_i^\top P (\by - \rE \by) \right| \geq \epsilon_2\right)
\leq 2 e^{-\epsilon_2^2/(2\sigma^2)} .
\]
Taking union bound for all $q_i \in Q$, we obtain
with probability exceeding
$1-2 (1+2/\epsilon_{1})^{k}e^{- \epsilon_2^2/2\sigma^2}$:
\[
\left|q_i^\top P (\by - \rE \by) \right| \leq \epsilon_2
\]
for all $i$.

Let $\beta=P(\by - \rE \by)/\|P(\by-\rE\by)\|_2$,
then there exists $i$ such that  $\|P\beta-P q_i\|_2 \leq \epsilon_1$.
We have
\begin{align*}
  \|P(\by-\rE\by)\|_2=&
  \beta^\top (\by-\rE\by) \\
  \leq& \|P \beta-P q_i\|_2 \|P(\by-\rE\by)\|_2 + |q_i^\top P(\by-\rE\by)| \\
  \leq& \epsilon_1\|P(\by-\rE\by)\|_2 + \epsilon_2 .
\end{align*}
Therefore
\[
\|P(\by-\rE\by)\|_2 \leq \epsilon_2/(1-\epsilon_1) .
\]
Let $\epsilon_1=2/15$,  and
$\eta=2 (1+2/\epsilon_{1})^{k}e^{- \epsilon_2^2/2\sigma^2}$,
we have
\[
\epsilon_2^2 = 2\sigma^2 [(4k+1) \ln 2 - \ln \eta] ,
\]
and thus
\[
\|P(\by-\rE\by)\|_2 \leq \frac{15}{13} \sigma \sqrt{2(4k+1) \ln 2 - 2\ln \eta} .
\]
This simplifies to the desired bound.

\section{Proof of Theorem~\ref{thm:rip}}

We use the following lemma from \cite{HuangZhang09:group}.
\begin{lemma}
  Suppose $X$ is generated according to Theorem~\ref{thm:rip}.
  For any fixed set $F \subset \cI$ with $|F|=k$ and $0<\delta<1$,
  we have with probability exceeding
  $1-3(1+8/\delta)^{k}e^{- n \delta^2/8}$:
  \begin{equation}
  (1 -\delta)\|\beta\|_{2}\leq\frac{1}{\sqrt{n}}\|X_F \beta\|_{2}\leq(1+\delta)\|\beta\|_{2} \label{eq:Phi-norm-bound-all}
  \end{equation}
  for all $\beta \in \mathbb{R}^k$.
  \label{lem:B3}
\end{lemma}

\subsection*{Proof of Theorem~\ref{thm:rip}}

Since $\cl(F)$ is a coding length, we have
\begin{align*}
\sum_{F: |F|+\cl(F) \leq s } (1+8/\delta)^{|F|}
\leq& \sum_{F: |F|+ \gamma \cl(F) \leq s } (1+8/\delta)^{|F|} \\
\leq& \sum_{F} (1+8/\delta)^{s-\gamma \cl(F)}
= (1+8/\delta)^s \sum_{F} 2^{-\cl(F)} \leq (1+8/\delta)^s ,
\end{align*}
where we let $\gamma=1/\log_2(1+8/\delta)$ in the above derivation.

For each $F$, we know from Lemma~\ref{lem:B3} that
for all $\beta$ such that $\Fr(\beta) \subset F$:
\[
(1 -\delta)\|\beta\|_{2}\leq\frac{1}{\sqrt{n}}
\|X \beta\|_{2}\leq(1+\delta)\|\beta\|_{2}
\]
with probability exceeding $1-3(1+8/\delta)^{|F|}e^{- n \delta^2/8}$.

We can thus take the union bound over $F: |F|+\cl(F) \leq s$,
which shows that with probability exceeding
\[
1-\sum_{F: |F|+\cl(F) \leq s} 3(1+8/\delta)^{|F|}e^{- n \delta^2/8} ,
\]
the structured RIP in Equation (\ref{eq:Group-RIP}) holds.
Since
\[
\sum_{F: |F|+\cl(F) \leq s} 3(1+8/\delta)^{|F|}e^{ - n \delta^2/8}
 \leq 3 (1+8/\delta)^s e^{- n \delta^2/8} \leq
e^{-t} ,
\]
we obtain the desired bound.

\section{Proof of Theorem~\ref{thm:struct-constr}
and Theorem~\ref{thm:struct-pen}}

\begin{lemma}
  Suppose that Assumption~\ref{assump:noise} is valid.
  For any fixed subset $F \subset \cI$,
  we have with probability $1-\eta$,
  $\forall \beta$ such that $\Fr(\beta) \subset F$, and
  $a>0$, we have
  \[
  \|X \beta - \rE \by\|_2^2
  \leq (1+a) [\|X \beta - \by\|_2^2 -  \|\by - \rE \by\|_2^2] +
  (2+a+a^{-1}) \sigma^2 [7.4 |F| + 4.7 \ln (4/\eta) ] .
  \]
  \label{lem:emp1}
\end{lemma}
\begin{proof}
  Let
  \[
  P_F= X_F (X_F^\top X_F)^{-1} X_F^\top
  \]
  be projection matrix to the subspaces generated by columns of $X_F$.

  Let $\tilde{a}=(I -P_F) \rE \by/\|(I -P_F)\rE \by\|_2$,
  $\delta_1=\|P_F(\by-\rE \by)\|_2$ and
  $\delta_2=|\tilde{a}^\top (\by - \rE \by)|$, we have
  \begin{align*}
&  \|X \beta-\rE\by\|_2^2 \\
=&
  \|X \beta - \by\|_2^2 - \|\by  - \rE \by\|_2^2 + 2 (\by - \rE \by)^\top (X \beta - \rE \by) \\
=&
  \|X \beta - \by\|_2^2 - \|\by  - \rE \by\|_2^2 + 2 (\by - \rE \by)^\top (X \beta - P_F \rE \by) - 2 \tilde{a}^\top (\by-\rE \by) \|(I-P_F) \rE \by\|_2 \\
= &
  \|X \beta - \by\|_2^2 - \|\by  - \rE \by\|_2^2 + 2 (\by - \rE \by)^\top P_F (X \beta - P_F \rE \by) - 2 \tilde{a}^\top (\by-\rE \by) \|(I-P_F) \rE \by\|_2 \\
\leq &
  \|X \beta - \by\|_2^2 - \|\by  - \rE \by\|_2^2 + 2 \delta_1  \|X \beta - P_F \rE \by\|_2 + 2 \delta_2 \|(I-P_F) \rE \by\|_2 \\
\leq &
  \|X \beta - \by\|_2^2 - \|\by  - \rE \by\|_2^2 + 2
\sqrt{\delta_1^2+\delta_2^2}
\sqrt{\|X \beta - P_F \rE \by\|_2^2 + \|(I-P_F) \rE \by\|_2^2} \\
= &
  \|X \beta - \by\|_2^2 - \|\by  - \rE \by\|_2^2 + 2
\sqrt{\delta_1^2+\delta_2^2} \|X \beta - \rE \by\|_2 .
\end{align*}
Note that in the above derivation,
we have used the fact that $P_F X \beta= X \beta$, and
$\|X \beta - P_F \rE \by\|_2^2 + \|(I-P_F) \rE \by\|_2^2= \|X \beta - \rE \by\|_2^2$.

Now, by solving the above inequality, we obtain
\begin{align*}
\|X \beta-\rE\by\|_2^2
\leq &
\left[\sqrt{\|X \beta - \by\|_2^2 - \|\by  - \rE \by\|_2^2 + \delta_1^2+\delta_2^2} + \sqrt{\delta_1^2+\delta_2^2} \right]^2 \\
\leq &
(1+a)
[\|X \beta - \by\|_2^2 - \|\by  - \rE \by\|_2^2] + (2+a+1/a)
(\delta_1^2 + \delta_2^2) .
\end{align*}
The desired bound now follows easily
from Proposition~\ref{prop:noise} and
Lemma~\ref{lem:sub-Gaussian}, where we know that with probability
$1-\eta/2$,
\[
\delta_1^2 =(\by - \rE \by)^\top P_F (\by - \rE \by)
\leq \sigma^2 (7.4 |F| + 2.7 \ln (4/\eta)) ,
\]
and with probability $1-\eta/2$,
\[
\delta_2^2
=|\tilde{a}^\top (\by - \rE \by)|^2 \leq 2\sigma^2 \ln(4/\eta) .
\]
We obtain the desired result by
substituting the above two estimates and simplify.
\end{proof}

\begin{lemma}
  Suppose that Assumption~\ref{assump:noise} is valid.
  Then we have with probability $1-\eta$,
  $\forall \beta \in \R^p$ and $a>0$:
  \[
  \|X \beta - \rE \by\|_2^2
  \leq (1+a) \left[\|X \beta - \by\|_2^2 -  \|\by - \rE \by\|_2^2\right] +
  (2+a + 1/a) \sigma^2 [7.4  c(\beta) +  4.7 \ln (4/\eta) ] .
  \]
  \label{lem:emp2}
\end{lemma}
\begin{proof}
  Note that for each $F$,
  with probability $2^{-\cl(F)} \eta$, we obtain
  from Lemma~\ref{lem:emp1} that $\forall \Fr(\beta) \in F$,
  \[
  \|X \beta - \rE \by\|_2^2
  \leq (1+a) \left[\|X \beta - \by\|_2^2 -  \|\by - \rE \by\|_2^2\right] +
  (2+a + 1/a) \sigma^2 [7.4  (|F|+\cl(F)) +  4.7 \ln (4/\eta) ] .
  \]
  Since $\sum_{F \subset \cI, F \neq \emptyset} 2^{-\cl(F)} \eta \leq \eta$, the result follows
  from the union bound.
\end{proof}

\begin{lemma}
  Consider a fixed subset $\bar{F} \subset \cI$.
  Given any $\eta \in (0,1)$, we have with probability
  $1-\eta$:
  \[
  |\|X \bb - \by\|_2^2 - \|\by - \rE \by\|_2^2| \leq
  \|X \bb - \rE \by \|_2^2 + 2 \sigma \sqrt{2 \ln(2/\eta)} \|X \bb - \rE \by \|_2.
  \]
  \label{lem:emp3}
\end{lemma}
\begin{proof}
  Let $\tilde{a}=(X \bb - \rE \by)/\|X \bb -\rE \by\|_2$, we have
  \begin{align*}
    & |\|X \bb - \by\|_2^2 -  \|\by - \rE \by\|_2^2| \\
    = & |-2 (X \bb  -\rE \by )^\top (\by - \rE \by) + \|X\bb - \rE \by\|_2^2| \\
    \leq & 2
    \|X \bb - \rE \by \|_2 |\tilde{a}^\top (\by-\rE \by)|
   + \|\rE \by - X \bb \|_2^2 .
  \end{align*}
  The desired result now follows from Lemma~\ref{lem:sub-Gaussian}.
\end{proof}

\begin{lemma} \label{lem:struct-pen}
Suppose that Assumption~\ref{assump:noise} is valid.
Consider any fixed target $\bb \in \R^p$.
Then with probability exceeding $1-\eta$,
for all $\lambda \geq 0, \epsilon \geq 0, \hb \in \R^p$ such that:
$\hQ(\hb) + \lambda c(\hb) \leq
\hQ(\bb) + \lambda c(\bb) + \epsilon$,
and for all $a>0$, we have
\begin{align*}
\|X \hb - \rE \by\|_2^2 \leq & (1+a)
  [ \|X \bb - \rE \by \|_2^2 + 2 \sigma \sqrt{2 \ln(6/\eta)} \|X \bb - \rE \by \|_2 ] \\
& \quad +
  (1+a) \lambda c(\bb) +   a' c(\hb) + b' \ln (6/\eta) + (1+a) \epsilon ,
\end{align*}
where $a'=7.4 (2+a+a^{-1}) \sigma^2 - (1+a) \lambda$ and
$b'=4.7 \sigma^2 (2+a+a^{-1})$.
Moreover, if the coding scheme $c(\cdot)$ is sub-additive, then
\[
n \rho_-(c(\hb)+c(\bb)) \|\hb- \bb \|_2^2
\leq 10 \|X \bb- \rE\by\|_2^2 +
2.5 \lambda c(\bb) +
( 37 \sigma^2 - 2.5 \lambda) c(\hb)
+ 29 \sigma^2 \ln (6/\eta) + 2.5 \epsilon .
\]
\end{lemma}
\begin{proof}
We obtain from the union bound of
Lemma~\ref{lem:emp2} (with probability $1-\eta/3$) and
Lemma~\ref{lem:emp3} (with probability $1-2\eta/3$) that
with probability $1-\eta$:
\begin{align*}
  &\|X \hb - \rE \by\|_2^2 \\
  \leq & (1+a) \left[\|X\hb - \by\|_2^2 -  \|\by - \rE \by\|_2^2\right] +
  (2+a+a^{-1}) [7.4\sigma^2 c(\hb) + 4.7 \sigma^2 \ln (6/\eta)] \\
  \leq & (1+a) \left[\|X\bb - \by\|_2^2 -  \|\by - \rE \by\|_2^2
    + \lambda c(\bb) + \epsilon \right] +
  a' c(\hb) + b' \ln (6/\eta) \\
  \leq & (1+a)
  [ \|X \bb - \rE \by \|_2^2 + 2 \sigma \sqrt{2 \ln(6/\eta)} \|X \bb - \rE \by \|_2 ] +  (1+a) \lambda c(\bb) + a' c(\hb)\\
  &\qquad   + b' \ln (6/\eta) + (1+a) \epsilon .
\end{align*}
This proves the first claim of the theorem.

The first claim with $a=1$ implies that
\begin{align*}
& \|X \hb- X \bb\|_2^2
\leq [\|X \hb- \rE \by \|_2 + \|X \bb- \rE\by\|_2 ]^2 \\
\leq& 1.25 \|X \hb- \rE \by \|_2^2 + 5 \|X \bb- \rE\by\|_2^2 \\
\leq& 7.5 \|X \bb- \rE\by\|_2^2 + 5 \sigma \sqrt{2\ln(6/\eta)} \|X \bb- \rE\by\|_2
+ 2.5 \lambda c(\bb) +
1.25 (29.6 \sigma^2 - 2 \lambda) c(\hb) \\
&\qquad + 1.25 \times 18.8 \sigma^2 \ln (6/\eta) + 2.5 \epsilon \\
\leq& 10 \|X \bb- \rE\by\|_2^2 +
2.5 \lambda c(\bb) +
( 37 \sigma^2 - 2.5 \lambda) c(\hb)
+ 29 \sigma^2 \ln (6/\eta) + 2.5 \epsilon .
\end{align*}
Since
$c(\hb-\bb) \leq c(\hb) + c(\bb)$,
we have
$\|X \hb- X \bb\|_2^2  \geq n \rho_-(c(\hb)+c(\bb)) \|\hb-\bb\|_2^2$.
This implies the second claim.
\end{proof}

\subsection*{Proof of Theorem~\ref{thm:struct-constr}}

We take $\lambda=0$ in Lemma~\ref{lem:struct-pen}, and obtain:
\begin{align*}
\|X \hb - \rE \by\|_2^2
\leq & (1+a)
  [ \|X \bb - \rE \by \|_2^2 + 2 \sigma \sqrt{2 \ln(6/\eta)} \|X \bb - \rE \by \|_2 ] \\
& \quad +
7.4 (2+a+a^{-1}) \sigma^2
c(\hb) + 4.7 \sigma^2 (2+a+a^{-1}) \ln (6/\eta) + (1+a) \epsilon \\
= &
  (\|X \bb - \rE \by \|_2 + \sigma \sqrt{2 \ln(6/\eta)})^2
+ 14.8 \sigma^2 c(\hb) + 7.4 \sigma^2 \ln(6/\eta) + \epsilon \\
& \quad +
a  [ (\|X \bb - \rE \by \|_2 +  \sigma \sqrt{2 \ln(6/\eta)})^2  + 7.4 \sigma^2 c(\hb) + 2.7 \sigma^2 \ln (6/\eta) + \epsilon ] \\
& \quad +
a^{-1} [ 7.4 \sigma^2 c(\hb) + 4.7 \sigma^2 \ln (6/\eta) ] .
\end{align*}
Now let $z=\|X \bb - \rE \by \|_2 + \sigma \sqrt{2 \ln(6/\eta)}$, and
we choose $a$ to minimize the right hand side as:
\begin{align*}
\|X \hb - \rE \by\|_2^2
\leq & z^2
+ 14.8 \sigma^2 c(\hb) + 7.4 \sigma^2 \ln(6/\eta) + \epsilon \\
& \quad +
2
[ z^2  + 7.4 \sigma^2 c(\hb) + 2.7 \sigma^2 \ln (6/\eta) + \epsilon ]^{1/2}
[ 7.4 \sigma^2 c(\hb) + 4.7 \sigma^2 \ln (6/\eta) ]^{1/2} \\
\leq &
[ (z^2  + 7.4 \sigma^2 c(\hb) + 2.7 \sigma^2 \ln (6/\eta) + \epsilon )^{1/2}
+ ( 7.4 \sigma^2 c(\hb) + 4.7 \sigma^2 \ln (6/\eta) )^{1/2} ]^2 \\
\leq &
[ z+ 2 (7.4 \sigma^2 c(\hb) + 4.7 \sigma^2 \ln (6/\eta) + \epsilon )^{1/2}]^2 .
\end{align*}
This proves the first inequality.
The second inequality follows directly from
Lemma~\ref{lem:struct-pen} with $\lambda=0$.

\subsection*{Proof of Theorem~\ref{thm:struct-pen}}

The desired bound is a direct consequence of Lemma~\ref{lem:struct-pen},
by noticing that
\[
2 \sigma \sqrt{2 \ln(6/\eta)} \|X \bb - \rE \by \|_2
\leq a \|X \bb - \rE \by \|_2^2 + a^{-1}
2 \sigma^2 \ln(6/\eta) ,
\]
$a' \leq 0$, and
\[
b'+ a^{-1} 2\sigma^2  \leq (10+5a+7a^{-1}) \sigma^2 .
\]

\section{Proof of Theorem~\ref{thm:greedy-conv1}
and Theorem~\ref{thm:greedy-conv2}}

The following lemma is an adaptation of a similar result in
\cite{Zhang08-foba} on greedy algorithms for standard sparsity.
\begin{lemma} \label{lem:one-step}
  Suppose the coding scheme is sub-additive. Consider any $\bb$, and a cover
  of $\bb$ by $\cB$:
  \[
  \Fr(\bb) \subset \bar{F}=\cup_{j=1}^b \bar{B}_j \quad (\bar{B}_j \in \cB) .
  \]
  Let $c(\bb,\cB)= \sum_{j=1}^b c(\bar{B}_j)$.
  Let $\rho_0 =\max_j \rho_+(\bar{B}_j)$.
  Then for all $F$ such that
  $c(\bar{B}_j \cup F) \geq c(F)$,
  \[
  \beta = \arg\min_{\beta' \in \R^p} \|X\beta'-\by\|_2^2 \quad \text{ subject to } \quad \Fr(\beta') \subset F ,
  \]
  and $\|X\beta-\by\|_2^2 \geq \|X\bb-\by\|_2^2$, we have
  \[
  \max_{j} \phi(\bar{B}_j)
  \geq \frac{\rho_-(F \cup \bar{F})}{\rho_0 c(\bb,\cB)}
  [\|X\beta-\by\|_2^2 - \|X\bb-\by\|_2^2] ,
  \]
  where as in (\ref{eq:phi}), we define
  \[
  \phi(B)=\frac{\|P_{B-F}(X\beta-\by)\|_2^2}
  {c(B \cup F)-c(F)} .
  \]
\end{lemma}
\begin{proof}
For all $\ell \in F$, $\|X \beta + \alpha X \be_\ell - \by\|_2^2$
achieves the minimum at $\alpha=0$ (where $\be_\ell$ is the vector
of zeros except for the $\ell$-th component, which is one).
This implies that
\[
\bx_\ell^\top (X \beta - \by)=0
\]
for all $\ell \in F$.
Therefore we have
\begin{align*}
 & (X \beta - \by)^\top \sum_{\ell \in \bar{F}-F} (\bb_\ell-\beta_\ell) \bx_\ell \\
=& (X \beta - \by)^\top \sum_{\ell\in \bar{F}\cup F} (\bb_\ell-\beta_\ell) \bx_\ell
= (X \beta - \by)^\top (X\bb-X\beta) \\
=& -\frac{1}{2} \|X (\bb -\beta)\|_2^2 + \frac{1}{2}\|X\bb-\by\|_2^2
-\frac{1}{2}\|X\beta-\by\|_2^2 .
\end{align*}
Now, let $\bar{B}_j' \subset \bar{B}_j-F$ be disjoint sets
such that $\cup_j \bar{B}_j' = \bar{F}-F$.
The above inequality leads to the following derivation $\forall \eta>0$:
\begin{align*}
&-\sum_j \phi(\bar{B}_j) (c(\bar{B}_j \cup F)-c(F)) \\
\leq &\sum_{j} \left[
\left\|X \beta + \eta \sum_{\ell \in \bar{B}_j'} (\bb_\ell-\beta_\ell) \bx_\ell  - \by
\right\|_2^2 - \|X \beta - \by\|_2^2 \right] \\
\leq& \eta^2 \sum_{\ell \in \bar{F}-F} (\bb_\ell-\beta_\ell)^2 \rho_0 n
+ 2 \eta (X \beta - \by)^\top \sum_{\ell \in \bar{F}-F} (\bb_\ell-\beta_\ell) \bx_\ell \\
\leq& \eta^2 \sum_{\ell \in \bar{F}-F} (\bb_\ell-\beta_\ell)^2 \rho_0 n
-\eta \|X (\bb -\beta)\|_2^2 + \eta \|X\bb-\by\|_2^2
-\eta \|X\beta-\by\|_2^2 .
\end{align*}
Note that we have used the fact that
$\|P_{B-F} (X \beta - \by)\|_2^2 \geq \|X \beta - \by\|_2^2-\|X \beta - \by +X \Delta \beta\|_2^2$ for all $\Delta \beta$ such that $\Fr(\Delta \beta) \subset B-F$.
By optimizing over $\eta$, we obtain
\begin{align*}
\max_j \phi(\bar{B}_j) \sum_j c(\bar{B}_j)
\geq&\sum_j \phi(\bar{B}_j) (c(\bar{B}_j \cup F)-c(F)) \\
\geq& \frac{[\|X (\bb -\beta)\|_2^2 + \|X\beta-\by\|_2^2 - \|X\bb-\by\|_2^2]^2}{4\sum_{\ell \in \bar{F}-F} (\bb_\ell-\beta_\ell)^2 \rho_0 n} \\
\geq& \frac{4 \|X (\bb -\beta)\|_2^2 [\|X\beta-\by\|_2^2 - \|X\bb-\by\|_2^2]}{4\sum_{\ell \in \bar{F}-F} (\bb_\ell-\beta_\ell)^2 \rho_0 n} \\
\geq& \frac{\rho_-(F \cup \bar{F})}{\rho_0}
 [\|X\beta-\by\|_2^2 - \|X\bb-\by\|_2^2] .
\end{align*}
This leads to the desired bound.
\end{proof}

\subsection*{Proof of Theorem~\ref{thm:greedy-conv1}}

Let
\[
\gamma'=   \frac{\gamma \rho_-(s + c(\bar{F}))}{\rho_0(\cB) c(\bb,\cB)} .
\]
By Lemma~\ref{lem:one-step}, we have at any step $k > 0$:
\[
\|X \beta^{(k-1)} - \by\|_2^2 - \|X \beta^{(k)} - \by\|_2^2
\geq \gamma'
[\|X\beta^{(k-1)}-\by\|_2^2 - \|X\bb-\by\|_2^2] (c(\beta^{(k)})-c(\beta^{(k-1)}) ,
\]
which implies that
\[
\max[0,\|X \beta^{(k)} - \by\|_2^2 - \|X\bb-\by\|_2^2]
\leq \max[0,\|X\beta^{(k-1)}-\by\|_2^2 - \|X\bb-\by\|_2^2]
e^{-  \gamma' (c(\beta^{(k)})-c(\beta^{(k-1)})} .
\]
Therefore at stopping, we have
  \begin{align*}
  &\|X \beta^{(k)} - \by\|_2^2 - \|X\bb-\by\|_2^2\\
  \leq& [\|\by\|_2^2 - \|X\bb-\by\|_2^2]
  e^{-  \gamma' c(\beta^{(k)})}   \\
  \leq& [\|\by\|_2^2 - \|X\bb-\by\|_2^2]
  e^{-  \gamma' s}  \leq \epsilon .
\end{align*}
This proves the theorem.

\subsection*{Proof of Theorem~\ref{thm:greedy-conv2}}

For simplicity, let $f_j=\hQ(\bb_j)$.
For each $k$, let $j_k$ be the largest $j$ such that
\[
\hQ(\beta^{(k)})
\geq f_j + f_j - f_0 + \epsilon .
\]
Let $\gamma'=(\gamma \min_j \rho_-(s + c(\bb_j)))/(\rho_0(\cB)  c(\bb_0,\cB))$.

We prove by contradiction. Suppose that
the theorem does not hold, then for all $k$
before stopping, we have $j_k \geq 0$.

For each $k>0$ before stopping,
if $j_k=j_{k-1}=j$, then we have from Lemma~\ref{lem:one-step}
(with $\bb=\bb_j$)
\[
c(\beta^{(k)}) \leq
c(\beta^{(k-1)}) + \gamma'^{-1} 2^{-j}
\ln \frac{\|X \beta^{(k-1)} - \by\|_2^2 - f_j}{\|X \beta^{(k)} - \by\|_2^2 - f_j} .
\]
Therefore for each $j \geq 0$, we have:
\[
\sum_{k: j_k=j_{k-1}=j} [c(\beta^{(k)}) -c(\beta^{(k-1)})]
\leq \gamma'^{-1} 2^{-j}
\ln \frac{2(f_{j+1} - f_0 +\epsilon)}{f_j - f_0 +\epsilon} .
\]

Moreover, for each $j \geq 0$,
Lemma~\ref{lem:one-step} (with $\bb=\bb_j$) implies that
\[
\sum_{k: j_k=j, j_{k-1}>j} [c(\beta^{(k)}) -c(\beta^{(k-1)})]
\leq \gamma'^{-1} 2^{-j} .
\]
Therefore we have
\[
\sum_{k: j_k=j} [c(\beta^{(k)}) -c(\beta^{(k-1)})]
\leq \gamma'^{-1} 2^{-j}
\left[1.7 +
\ln \frac{f_{j+1} - f_0 +\epsilon}{f_j - f_0+\epsilon} \right] .
\]
Now by summing over $j \geq 0$, we have
\[
c(\beta^{(k)})
\leq 3.4 \gamma'^{-1}
+\gamma'^{-1}
\sum_{j=0}^\infty 2^{-j}\ln \frac{f_{j+1} - f_0 +\epsilon}{f_j - f_0 +\epsilon}  \leq s .
\]
This is a contradiction because we know at stopping, we should have
$c(\beta^{(k)}) >s$.

\section{Proof of Corollary~\ref{cor:greedy-compressible}}
Given $s'$, we consider $f_j=\min_{\ell \geq j} \hQ(\bb(s'/2^\ell))$.
We may assume that
$f_0$ is achieved with $\ell_0=0$.
Note that by
Lemma~\ref{lem:emp3}, we have with probability $1-2^{-j-1} \eta$:
\begin{align*}
|\hQ(\bb(s'/2^j)) -\|\by - \rE \by\|_2^2|
\leq& 2 \|X \bb(s'/2^j) - \rE \by\|_2^2 + 2 \sigma^2 [j+1 + \ln(2/\eta)] \\
\leq& 2 a n 2^{qj}/s'^q + 2\sigma^2 [j+1 + \ln(2/\eta)] .
\end{align*}
This means the above inequality holds for all $j$ with probability $1-\eta$.
Now, by taking $\epsilon= 2 a n /s'^q + 2 \sigma^2 [\ln (2/\eta) +1]$ in Theorem~\ref{thm:greedy-conv2}, we obtain
\begin{align*}
 \sum_{j=0}^\infty 2^{-j} \ln \frac{f_{j+1}-f_0+\epsilon}{f_j-f_0+\epsilon}
\leq&
\sum_{j=\ell_0}^\infty 2^{-j} \ln (1+ (f_{j+1}-f_0)/\epsilon) \\
\leq&
\sum_{j=\ell_0}^\infty 2^{-j} \ln (2 + 2 (j+2^{q(j+1)})) \\
\leq&
\sum_{j=\ell_0}^\infty 2^{-j} (\ln 2+ 1 + j + q (j+1) \ln 2 ) \leq 2 + 4 (1+ q \ln 2) ,
\end{align*}
where we have used the simple inequality $\ln(\alpha+\beta) \leq \alpha  +\ln(\beta)$
when $\alpha,\beta \geq 1$.
Therefore,
\begin{align*}
s \geq &
\frac{\rho_0(\cB) s'}{\gamma \min_{u \leq s'} \rho_-(s + c(\bb(u)))}
(10 + 3 q) \\
\geq&
\frac{\rho_0(\cB) s'}{\gamma \min_{u \leq s'} \rho_-(s + c(\bb(u)))}
\left[ 3.4 + \sum_{j=0}^\infty 2^{-j}
  \ln \frac{f_{j+1}-f_0+\epsilon}{f_j-f_0+\epsilon}
\right] .
\end{align*}
This means that Theorem~\ref{thm:greedy-conv2} can be applied
to obtain the desired bound.

\end{document}